\documentclass[referee]{aa}
\usepackage{graphicx}
\usepackage{natbib}
\usepackage{txfonts}

\begin{document}

\title{Formation and evolution of the Galactic bulge: constraints
from stellar abundances}

%\altaffiltext{1}{
%Some of the data discussed here were obtained
%at the W.M. Keck Observatory, which is operated as a scientific partnership
%among the California Institute of Technology, the University of California,
%and the National Aeronautics and Space Administration.
%The Observatory was made possible by the generous financial support of the
%W.M. Keck Foundation.}}

\author {S.K. Ballero\inst{1}
\thanks {email to: ballero@oats.inaf.it}
\and  F. Matteucci\inst{1, 2}
\and  L. Origlia\inst{3}
\and  R.M. Rich \inst{4}}

\institute{ Dipartimento di Astronomia, Universit\'a di Trieste, via G.B. 
Tiepolo 11, I-34131, Trieste, Italy  
\and  INAF- Osservatorio Astronomico di Trieste, via G.B. Tiepolo 11, 
I-34131, Trieste, Italy
\and INAF-Osservatorio Astronomico di Bologna, Via G. Ranzani 1,  I-40127 
Bologna, Italy
\and  Department of Physics and Astronomy, UCLA, 430 Portola Plaza,
Box 951547, Los Angeles, CA 90095-1547} 

\date{Received xxxx / Accepted xxxx}

\abstract{}
{To compute the chemical evolution of the Galactic
bulge in the context of an inside-out model for the formation of the
Milky Way. The model contains updated stellar yields from massive stars.
The main purpose of the paper  is to compare the predictions of this
model with new observations of chemical abundance ratios and metallicity
distributions in order to put constraints on the formation and evolution 
of the bulge.} 
{We computed the evolution of several $\alpha$-elements and Fe and performed 
several tests by varying different parameters such as star formation 
efficiency,
slope of the initial mass function and infall timescale. We also tested the 
effect of adopting a primary nitrogen contribution from massive stars.} 
{The [$\alpha$/Fe] abundance ratios in the Bulge are predicted to be
  supersolar for a very large range in [Fe/H], each element having a
  different slope. These predictions are in very good agreement with
  most recent accurate abundance determinations. We also find a good
  fit of the most recent Bulge stellar metallicity distributions.}  
{We conclude that the Bulge formed on
a very short timescale (even though timescales much shorter than $\sim
0.1$ Gyr are excluded) with a quite high star formation efficiency of
$\nu \simeq 20$ Gyr$^{-1}$ and with an initial mass function more
skewed toward high masses (i.e. $x \leq 0.95$) than the solar
neighbourhood and rest of the disk. The results obtained here are more
robust than previous ones since they are based on very accurate
abundance measurements.}

\keywords{Galaxy: bulge, Galaxy: evolution, Galaxy: abundances}

\titlerunning{Bulge evolution}             

\maketitle

\authorrunning{Ballero  et al.}

\section{Introduction}

Galactic bulges are spheroidal systems that are found in the centre
of most spiral galaxies, and usually possess metallicity, photometric
and kinematic properties that separate them from the disk
components. 
Studying the evolution of the bulge of the Galaxy is of general
interest because of the broad similarity of its integrated light to
elliptical galaxies and other spiral bulges (e.g. Whitford, 1978;
Maraston et al., 2003). Bulges and ellipticals are located in the same
regions of the fundamental plane 
(Jablonka et al., 1996); the Galactic bulge has a wide range in [Fe/H]
(Rich, 1988; McWilliam \& Rich, 2004; Zoccali et al., 2003; Fulbright
et al., 2005) and enhancement of $\alpha-$elements (McWilliam \& Rich,
1994; McWilliam \& Rich, 2004; Rich \& Origlia, 2005); the age
dispersion in the bulge at distances greater than 300 pc from the
nucleus is observed to be narrow.
Terndrup (1988) first argued for a globular cluster-like nature to the
bulge population. Ortolani et al. (1995) compared the luminosity
function of a metal rich globular cluster with that of the bulge field
Baade's Window. Recently,  Kuijken \& Rich (2002) and Zoccali et
al. (2003) showed that when the bulge field is decontaminated from disk
foreground stars by proper motion cleaning or statistical subtraction,
the remaining population is indistinguishable 
from an old metal rich globular cluster. These studies form the basis
for a growing consensus that the bulge is old and that its formation 
timescale was relatively short ($\leq 1$ Gyr).
Very recently, medium- and high-resolution spectroscopy of bulge stars
have been performed (Ram\'{\i}rez et al. 2000; Fulbright et al., 2006a,
FMR06a), providing further support to a fast formation of the bulge.
 
Finally, a short formation timescale for the bulge is also
suggested on theoretical grounds by Elmegreen (1999), who argued that
the potential well of the Galactic bulge is too deep to allow
self-regulation and that most of the gas must have been converted into
stars within a few dynamical timescales.
Moreover, Sarajedini \& Jablonka (2005) suggest that, since the
differences in the metallicity distributions (MDs) of the Milky Way
and M31 halos find no correspondence in those of their bulges, the
bulk of the stars in the bulges of both galaxies must have been in
place before any accretion event that might have occurred in the halo
could have any influence on them, supporting a common scenario for the
formation of bulges. 

Wyse (1998) also showed that the MD in our Galaxy is not consistent
with a picture where the bulge is formed via accretion of satellites
(see Searle \& Zinn, 1978).
There is now additional evidence that the bulge is not formed by 
satellites similar to those observed at the present day in the halos
of the Milky Way and M31. 
McWilliam et al. (2003) find a systematically decreased
abundance of Mn in the Galactic bulge, compared to stars in the Sgr
dwarf spheroidal. So, even though Sgr does in principle reach high
metallicity, its detailed chemistry is different.

 The baseline model for the chemical evolution of the bulge might
  well be the one-zone model (Searle \& Zinn, 1978) with its
  instantaneous recycling and closed-box assumptions. Early numerical
  models (Arimoto \& Yoshii, 1986) produced the observed wide
  abundance range. 
  When supernova yields and detailed stellar lifetimes are
  incorporated, more extensive predictions become possible. 
  This is
  the case of the Galactic bulge model by Matteucci \& Brocato (1990,
  MB90) who predicted that the [$\alpha$/Fe] ratio for 
  some elements (O, Si and Mg) should be supersolar over almost the
  whole metallicity range, in analogy with the halo stars, as a
  consequence of assuming a fast bulge evolution which involved rapid
  gas enrichment in Fe mainly by Type II SNe.  
At that time, no data were available for detailed chemical abundances;
the predictions of MB90 were later confirmed for a few
$\alpha$-elements (Mg, Ti) by the observations of McWilliam \& Rich
(1994, MR94). 
%, whereas for other elements (e.g. Ca, Si) the observed trend at that
% time was different. 
% Other discrepancies regarding the Mg overabundance came from Sadler et
% al. (1996).
A few years later, Matteucci et al. (1999, MRM99) modeled the
behaviour of a larger set of abundance ratios, by means of a detailed 
chemical evolution model whose parameters were calibrated so that the
metallicity distribution observed by MR94 could be fit.
They predicted the evolution of several abundance
ratios that were meant to be confirmed or disproved by subsequent
observations, namely that $\alpha$-elements should in general be
overabundant with respect to Fe, but some (e.g. Si, Ca) less than
other (e.g. O, Mg), and that the [$^{12}$C/Fe] ratio should be solar
at all metallicities. 
They concluded that an IMF index flatter ($x=1.1-1.35$) than that of
the solar neighbourhood is needed for the MR94 metallicity
distribution to be reproduced, and that an evolution much faster than
that in the solar neighbourhood and faster than that of the halo (see
also Renzini, 1993) is necessary as well. 
Whatever secular processes may be at work with respect to the
formation and maintenance of the bar, the chemistry and stellar ages
require that the stars formed early and self-enriched rapidly.

Not all models of the bulge support these conclusions.
Samland et al. (1997) developed a self-consistent
chemo-dynamical model for the evolution of the Milky Way components
starting from a rotating protogalactic gas cloud in virial
equilibrium, that collapses owing to dissipative cloud-cloud
collisions. 
They found that self-regulation due to a bursting star formation and
subsequent injection of energy from Type II supernovae led to the
development of ``contrary flows'', i.e. alternate collapse and outflow
episodes in the bulge. 
This caused a prolonged star formation episode lasting over
$\sim4\times10^9$ yr. 
They included stellar nucleosynthesis of O, N and Fe, but claim that
gas outflows prevent any clear correlation between local star
formation rate and chemical enrichment. 
With their model, they could reproduce the oxygen gradient of H
{\scshape ii} regions in the equatorial plane of the Galactic disk and
the metallicity distribution of K giants in the bulge (Rich, 1988),
field stars in the halo and G dwarfs in the disk, but they did not
make predictions about the evolution of abundance ratios in the
bulge.

Moll\'a et al. (2000) proposed a multiphase model in the context of
the dissipative collapse scenario of the Eggen et al. (1962) picture. 
They suppose that the bulge formation occurred in two main infall
episodes, the first from the halo to the bulge, on a timescale
$\tau_H=0.7$ Gyr (longer than that proposed by MRM99), and the
second from the bulge to a so-called core population in the very
nuclear region of the Galaxy, on a timescale $\tau_B \gg \tau_H$. 
The three zones (halo, bulge, core) interact via supernova winds and
gas infall. 
They concluded that there is no need for accretion of external
material to reproduce the main properties of bulges and that the
analogy to ellipticals is not justified.
Because of their rather long timescale for the bulge formation, these
authors did not predict a noticeable difference in the trend of the
[$\alpha$/Fe] ratios but rather suggested that they behave more likely
as in the solar neighbourhood (contrary to the first indications of
$\alpha$-enhancement by MR94).

A more recent model was proposed by Costa et al. (2005), in which the
best fit to the observations relative to planetary nebulae (PNe) is
achieved by means of a double infall model. 
An initial fast (0.1 Gyr) collapse of primordial gas is followed by a
supernova-driven mass loss and then by a second, slower (2 Gyr) infall
episode, enriched by the material ejected by the bulge during the
first collapse. 
Costa et al. (2005) claim that the mass loss is necessary to reproduce
the abundance distribution observed in PNe, and because
the predicted abundances would otherwise be higher than observed.
With their model, they are able to reproduce the trend of [O/Fe]
abundance ratio observed by Pomp\'eia et al. (2003) and the data of
nitrogen versus oxygen abundance observed by Escudero \& Costa (2001)
and Escudero et al. (2004). It must be noted however that 
Pomp\'eia et al. (2003) obtained abundances for ``bulge-like'' dwarf
stars. This ``bulge-like'' population consists of old ($\sim 10-11$
Gyr), metal-rich nearby dwarfs with kinematics and metallicity
suggesting an inner disk or bulge origin and a mechanism of radial
migration, perhaps caused by the action of a Galactic bar. 
However, the birthplace of these stars is not certain, therefore we
decided to omit these data from our model discussions, preferring to
consider those stars for which membership in the present day bulge is
secure. 
Moreover, as we shall see, the use of nitrogen
abundance from PNe is questionable, since N is known to be also
synthesized by their progenitors and therefore it might not be the
pristine one. 

In this paper we want to test the hypothesis of a quick dissipational
collapse via the study of the evolution of the abundance ratios
coupled with considerations on the metallicity distribution. 
Ferreras et al. (2003) already tried to fit the stellar
  metallicity distributions of K giants of Sadler et al. (1996), Ibata
  \& Gilmore (1995) and Zoccali et al. (2003), which are pertinent to
  different bulge fields, by means of a model of star
  formation and chemical evolution. 
  Their model assumes a Schmidt law
  similar to that of the disk, and simple recipes with a few
  parameters controlling infall and continuous outflow of gas. They
  explore a large range in parameter space and conclude that
  timescales longer than $\sim1$ Gyr must be excluded at the 90\%
  confidence level, regardless of which field is being considered. 
We want to show that abundance ratios can provide
an independent constraint for the bulge formation scenario since they
differ depending on the star formation history (Matteucci, 2000).

We focus our attention on the evolution of the $\alpha$-elements,
carbon and nitrogen as a function of metallicity. 
New data have lately become available for the abundance ratios of
these nuclids: $\alpha$-elements in particular are of paramount
importance in probing the star formation timescale, since the
signature of a very short burst of star formation must result into
an $\alpha$-enhancement with respect to iron, whereas $\sim$solar
abundances would imply that Type Ia Supernovae (SNe) had time to
pollute the ISM with iron-rich ejecta (see e.g. Matteucci \& Greggio,
1986). 
Moreover, we also analysed the evolution of nitrogen and
studied the possibility of primary N being produced by all massive
stars at any metallicity, as described in Matteucci (1986). This
seems to provide a good fit to [N/O] and [N/Fe] in the solar vicinity
(Chiappini et al., 2005; Ballero et al. 2006; see also Meynet \&
Maeder, 2002).
The present paper does not primarily mean to point at a particular
model as the ``best model'', but rather has the aim of showing which
are the possible effects of varying the above-mentioned parameters; in
fact, since the results for the MD of bulge giants are still
preliminary, and high-resolution data are awaited, it is premature to
draw firm conclusions on which combination of parameters provides the
best fit. However, we can quite safely restrict the ensemble of
plausible models by means of the present analysis.

The paper is organized as follows: \S 2 describes the chemical
evolution model, \S 3 provides a review of the data employed, \S 4
shows the model results regarding the supernova rates, the chemical
evolution of abundance ratios, the MD of G, K and M giants in 
the bulge and in \S 5 we draw our conclusions.

\section{The chemical evolution models}
\label{sec:model}

The adopted basic chemical evolution model closely follows that
in MRM99. 
The main assumption is that the Galactic bulge formed with the
fast collapse of primordial gas (the same gas out of which the
halo was formed) accumulating in the centre of our Galaxy. 
We recall the fundamental ingredients of this model:
\begin{itemize}
\item[-] Instantaneous mixing approximation: the gas over the whole
  bulge is homogeneous and well mixed at any time.
\item[-] Star formation rate (SFR) parametrized as follows:
\begin{equation}
\psi(r,t) =
\nu G^k(r,t)
\end{equation}
where $\nu$ is the star formation efficiency (i.e. the inverse of the
timescale of star formation) in the bulge, $k = 1$ is chosen to
recover the star formation law employed in models of spheroids
(e.g. by Matteucci, 1992) and $G(r,t) =
\sigma_{gas}(r,t)/\sigma(r,t_G)$ is the normalized gas surface mass 
density (where $\sigma_{gas}(r,t)$ is the gas surface mass density
and $\sigma(r,t_G)$ is the surface gas density of the bulge at the
present time $t_G=13.7$ Gyr). 
We tested also different values for $k$ (such as $1.5$, which is
  the disk value) and the result do not differ much. The main
  difference with the solar neighbourhood resides in the higher $\nu$
  parameter for the bulge. 
\item[-] We did not adopt a thresold surface gas density for the onset
  of star formation such as that proposed by Kennicutt (1998) for the
  solar neighbourhood, since it is derived for self-regulated disks
  and there seems to be no reason for it to hold also in early galaxy
  evolutionary conditions and in bulges. However, we also checked that
  adopting a threshold of $4~M_{\odot}$pc$^{-2}$ such as that proposed
  by Elmegreen (1999) does not change our results, since a wind
  (see below) develops much before such a low gas density is reached.
\item[-] The initial mass function (IMF) is expressed as a power
  law with index $x$:
\begin{equation}
\phi(m) \propto m^{-(1+x)}
\end{equation}
within the mass range $0.1-100M_{\odot}$.
\item[-] The gas forming the bulge has a primordial chemical
  composition and accretes at a rate given by:
\begin{equation}
\dot{G}(r,t)_{inf}=\frac{A(r)}{\sigma(r,t_G)}e^{-t/\tau}
\end{equation}
where $\tau$ is an appropriate collapse timescale and $A(r)$ is
constrained by the requirement of reproducing the current total
surface mass density in the Galactic bulge. 
Actually, we should use the halo chemical composition for the
infalling gas, but it can be demonstrated that unless very high
$\alpha$-enhancements are adopted, the results are essentially the
same. In principle, a slightly enriched infall could improve the
  fit of the MDs (see later) but this is not likely to have
  significant effects for realistic levels of enrichment. 
\noindent\item[-] The instantaneous recycling approximation is relaxed;
  stellar lifetimes are taken into account in detail following the
  prescriptions of Kodama (1997).
\noindent\item[-] Detailed nucleosynthesis prescriptions are taken from Fran\c
  cois et al. (2004), who made use of widely adopted stellar yields
  and compared the results obtained by including these yields in
  the model from Chiappini et al. (2003a) with the observational data,
  with the aim of constraining the stellar nucleosynthesis. 
  Namely, for low- and intermediate-mass ($0.8 - 8M_{\odot}$) stars,
  which produce $^{12}$C, N and heavy $s$-elements, yields are taken
  from the standard model of Van den Hoek \& Groenewegen (1997) as a
  function of the initial stellar metallicity. Concerning massive
  stars ($M>10M_{\odot}$), in order to best fit the data in the solar 
  neighbourhood, when adopting Woosley \& Weaver (1995)
  \footnote{
  The models of Woosley \& Weaver (1995) do not include mass loss,
  which might not be realistic for non-zero metallicity stars. 
  However, it was shown (Bressan et al., 1981; Bertelli et al., 1990)
  that the combined effects of mass loss and overshooting
  compensate each other, in the sense that the resulting He-core mass
  (where all the heavy elements are produced) remains approximately
  the same as that in models without mass loss and overshooting. 
  Additionally, nucleosynthesis with mass loss and rotation has so
  far been computed only for a few elements (e.g. Maeder et al.,
  2005).
  } 
  yields, Fran\c cois et al. (2004) found that O yields should be
  adopted as a function of initial metallicity, Mg yields should be
  increased in stars with masses $11-20M_{\odot}$ and decreased in
  stars larger than $20M_{\odot}$, and that Si yields should be
  slightly increased  in stars above $40M_{\odot}$; we use their
  constraints on the stellar nucleosynthesis to test whether the same 
  prescriptions give good results for the Galactic bulge.
  Yields in the mass range $40-100M_{\odot}$ were not computed by
  Woosley and Weaver (1995), therefore one has to extrapolate them
  for chemical evolution purposes
  \footnote{The authors are
  aware that the extrapolation process is problematic. However, the
  behaviour above this mass is not clear, since a supernova explosion
  may occur with a large amount of fallback. Moreover, it was shown
  (Fran\c cois et al., 2004) that it is impossible to reproduce the
  observations at low metallicities in the solar neighbourhood if no
  contribution from stars in this mass range is considered.
  }.
\noindent\item[-] The Type Ia SN rate was computed according to Greggio \&
  Renzini (1983) and Matteucci \& Recchi (2001). 
  Yields are taken from Iwamoto et al. (1999) which is an updated
  version of model W7 (single degenerate) from Nomoto et al. (1984).
  These supernovae are the main contributors of Fe and produce small
  amounts of light elements; they also contribute to some extent to
  the enrichment in Si and Ca.
\noindent\item[-] Contrary to MRM99, we introduced the treatment of 
  a supernova-driven galactic wind in analogy with ellipticals
  (e.g. Matteucci, 1994). The bulge lies in the potential well of the
  Galactic disk and very massive dark halo that provides a high
  binding energy. Therefore, previous chemical evolution models
  supposed that a wind should not develop and moreover, the
  occurrence of winds did not seem to be suitable to reproduce the
  metallicity distribution of the Milky Way components (Tosi et
  al. 1998), especially for the disk.
  Moreover, Elmegreen (1999) sustained that the bulge potential well
  is too deep to allow for self-regulation and that the gas should be
  converted into stars in only a few dynamical timescales. 
  However, this scenario was not tested quantitatively.
  We therefore supposed that the bulge is bathed in a dark matter
  halo of mass 100 times greater than that of the bulge itself
  (i.e. $M_{dark} = 2 \times 10^{12} M_{\odot}$) and with an effective
  radius $r_{dark} = 100r_e = 200$ kpc, where $r_e$ is the
  effective radius of the bulge (S\'ersic) mass distribution.
  To compute the gas binding energy   $E_{b,gas}(t)$  we have
  followed Matteucci (1992)  who adopted the formulation of 
  Bertin et al. (1992). They analysed the properties of a family of
  self-consistent spherical two-component 
  models of elliptical galaxies, where the luminous mass is embedded
  in massive and diffuse dark halos, and in this
  context they computed  binding energy of the gas.
    A more refined treatment of the Galactic potential well would take
  into account also the contribution of the disk; however, it is
  easy to show that the main contributors to the bulge potential well
  are the bulge itself and the dark matter halo. 
  The condition for the onset of the galactic wind is:
  \begin{equation}
    E_{th,SN}(t_{GW})=E_{b,gas}(t_{GW})
  \end{equation}
  \label{eq1}
  where  $E_{th,SN}(t)$ is the thermal energy of the gas at the time $t$
  owing to the energy deposited by SNe (II and Ia) (see Matteucci 1992
  and Bradamante et al. 1998).        
  At the specific time $t_{GW}$ (the time for the occurrence of a
  galactic wind), all the remaining gas is expeled from the bulge, and 
  both star formation and gas infall cease.
  However, in all cases, the galactic wind occurs when most of the gas
  has already been converted into stars, and its effect on chemical
  evolution is negligible. Therefore, we shall limit ourselves to
  switching off star formation and infall when $t_{GW}$ is reached.
  \item We suppose that feedback from the central black hole is
  negligible. We shall test this hypothesis quantitatively in a
  forthcoming paper.
  \end{itemize}

As a fiducial model, we adopt the one with the following reference
parameters: $\nu=20$ Gyr$^{-1}$, collapse timescale $\tau=0.1$ Gyr and
a two-slope IMF with index $x=0.33$ for $M < 1 M_{\odot}$ and $x=0.95$
for $M > 1 M_{\odot}$ (Matteucci \& Tornamb\`e, 1987). 
The choice of such a flat IMF for the lowest-mass stars is motivated
by the Zoccali et al. (2000) work, who measured the luminosity
function of lower main-sequence bulge stars and derived the mass
function, which was found to be consistent with a power-law of index
$0.33 \pm 0.07$. 
The IMF index for intermediate-mass and massive stars which is 
slightly flatter than that adopted by MRM99 in order to reproduce the
MD of bulge stars from Zoccali et al. (2003, Z03) and Fulbright et al.
(2006a, FMR06a; see \S\ \ref{sec:data} and \ref{subs:GKM} for details)
instead ot that from MR94.
We explored a number of other possibilities by varying the model
parameters in the following way:
\begin{itemize} 
\item[-] Star formation efficiency: $\nu$ from 2 to 200 Gyr$^{-1}$;
\item[-] For the IMF above $1 M_{\odot}$, we have considered the cases
  suggested by Zoccali et al. (2000, Z00) in their \S 8.3, i.e. their
  case 1 (hereafter Z00-1) with $x=0.33$ in the whole range of masses, 
  their case 3 (Z00-3) for which $x=1.35$ for $M > 1M_{\odot}$
  (Salpeter, 1955) and their case 4 (Z00-4) in which $x=1.35$ for
  $1M_{\odot} < M < 2M_{\odot}$ and $x=1.7$ for $M > 2M_{\odot}$
  (Scalo, 1986). 
  Our reference model corresponds to their case 2 with $x=0.95$ for $M
  > 1M_{\odot}$
  \footnote{Actually, Z00 selected as ``IMF 2'' the one with $x=1$ for
  $M > 1M_{\odot}$, but we performed calculations with $x=0.95$, which
  is very similar, for comparison to the IMF with $x=0.95$ of
  (Matteucci \& Tornamb\`e, 1987).},
  therefore we call it Z00-2. 
  We recall that the fraction $A$ of binary systems giving rise to
  Type Ia SNe is a function of the adopted IMF (see Matteucci \&
  Greggio, 1986). 
  Owing to the lack of information concerning the Type Ia SN rate in
  the bulge, we calibrate such a fraction in order to reproduce the
  Mannucci et al. (2005) estimate of the SN rate of an elliptical
  galaxy of the same mass.
\item[-] Infall timescale: $\tau$ from 0.01 to 0.7 Gyr; the latest
  hypothesis follows the suggestion by Moll\'a et al. (2000) who
  assume a slower timescale for the formation of the bulge. We refer to
  the timescale $\tau_H$ which they chose for the gas collapse from
  halo to bulge.
\end{itemize}

\begin{table}
\centering
\footnotesize
\begin{tabular}{llll}
\hline
\hline
Model name/specification & 
$x$ ($M > 1 M_{\odot}$) &
$\nu$ (Gyr$^{-1}$) & 
$\tau$ (Gyr) \\
\hline
\hline
Fiducial model (Z00-2) & 0.95 & \phantom{0}20.0 & 0.1 \\
\hline
Z00-1 & 0.33 & \phantom{0}20.0 & 0.1 \\ 
Z00-3 & 1.35 & \phantom{0}20.0 & 0.1 \\  
Z00-4 & Scalo* & \phantom{0}20.0 & 0.1 \\
\hline
$\nu=2$ Gyr$^{-1}$ & 0.95 & \phantom{00}2.0 & 0.1 \\ 
$\nu=200$ Gyr$^{-1}$ & 0.95 & 200.0 & 0.1 \\ 
\hline
$\tau=0.01$ Gyr & 0.95 & \phantom{0}20.0 & 0.01 \\ 
$\tau=0.7$  Gyr & 0.95 & \phantom{0}20.0 & 0.7 \\ 
\hline
S1    & 0.33 & 200.0 & 0.01 \\ 
S2    & 0.95 & 200.0 & 0.01 \\ 
S3    & 0.33 & \phantom{00}2.0 & 0.7 \\ 
S4    & 0.95 & \phantom{00}2.0 & 0.7 \\
S5    & 0.33 & \phantom{00}0.5 & 1.25 \\
\hline
\hline
\end{tabular}
\caption{Features of the examined models.}
\label{tab:tab1}
\end{table}

Table \ref{tab:tab1} summarizes the features of the considered
models.\footnote{In the first column are indicated the names of the
  models, or their distinctive characteristic when no name was
  specified; in the second, third and fourth column are shown the IMF
  index, star formation efficiency and gas infall timescale respectively
  (* ``Scalo'' means $x=1.35$ for $1 <
  M/M_{\odot} <2$ and $x=1.7$ for $M > 2 M_{\odot}$).}

\section{The data}
\label{sec:data}

We now turn to the datasets against which the models will be
compared. These are re-normalized to the solar abundances of
Grevesse \& Sauval (1998) so that an artificial dispersion associated
with the adoption of different solar abundance values is corrected for. 
%In the figures
%regarding the evolution of element abundance ratios, for the sake of
%clarity errorbars are only shown in the [$\alpha$/Fe] plots, but they
%will be however indicated in the following.

\subsubsection*{Metallicity distribution}

Data for the [Fe/H] distribution of red giant branch (RGB) and
asymptotic giant branch (AGB) stars in the bulge were taken from
Z03, who provided photometric determination of metallicities for 503
bulge stars.
By combining near-infrared data from the 2MASS survey, from the SOFI
imager at ESO NTT and the NICMOS camera on HST, plus optical images
taken with the WFI at ESO/MPG 2.2m telescope within the EIS PRE-FLAMES
survey, they constructed a disk-decontaminated ($M_K$,$V-K$)
color-magnitude diagram (CMD) of the bulge stellar population with
very large statistics and small photometric errors, which was compared
with the analytical RGB templates in order to derive the MD. 
The advantage of this approach is that it allows determinations of
metallicities of a great number of stars, although the relationship
between the position in the CMD and the metallicity can be somewhat
uncertain.  
%, and that it
%avoids the very uncertain conversion from spectral indices to
%metallicities which is used in low-resolution spectroscopy. 
%However, it is not the most accurate way to measure metallicities
%since 

Since the template RGBs are on the [M/H] scale (where M stands for the
total metal abundance), in order to obtain the [Fe/H] distribution the
$\alpha$-enhancement contribution was subtracted in the following~way
(Zoccali, private communication):

\begin{equation}
\mbox{[Fe/H]}=
\left\{
\begin{array}{ll}
\mbox{[M/H]}-0.14 & \mbox{ if } \mbox{[M/H]}>-0.86 \\
\mbox{[M/H]}-0.21 & \mbox{ if } \mbox{[M/H]}<-0.86 
\end{array}
\right.
\end{equation}

This relation assumes that the $\alpha$-elements in the bulge follow
the abundance trends of globular clusters in the halo.
They found that the resulting MD contained somewhat less metal-poor
stars relative to a closed-box gas exhaustion model and that the
G-dwarf problem (i.e. the deficit of metal-poor stars relative to a
simple model) may affect the Galactic bulge even though less severely
than in the solar neighbourhood (see e.g. Hou et al., 1998).

\begin{figure}
\centering
\includegraphics[width=.5\textwidth]{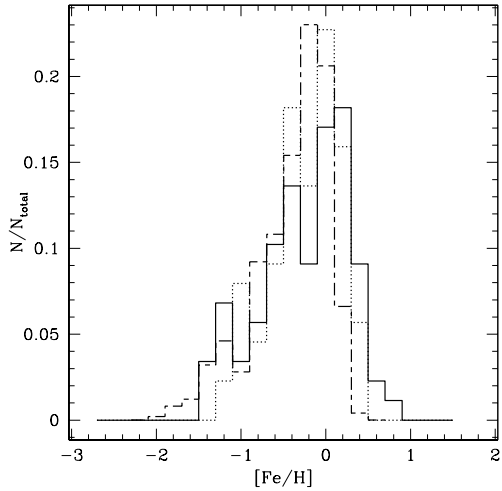}
\caption{Comparison between the bulge MD derived by Z03 (solid
  histogram), FMR06a (dashed histogram) and that derived with the
  spectroscopic survey of MR94 (dotted histogram).} 
\label{fig:fig01}
\end{figure}

In figure \ref{fig:fig01} we compare the photometric MD derived by
Z03 to the spectroscopic one of MR94. 
The two distributions are broadly consistent, with a somewhat
less pronounced supersolar [Fe/H] tail and a slightly sharper
  peak in the photometric case.  
However, the position of the high [Fe/H] cutoff for the photometric MD
is wholly dependent on the metallicity assigned to the only two template
clusters, namely NGC 6528 and NGC 6553 available in the
high-metallicity domain. 
%If such metallicity were $\sim0.2$ dex higher, 
%the discrepancy would disappear, i.e. the photometric MD would be
%stretched and the peak at [Fe/H]$\sim -0.2$ would be reduced.
Nevertheless, there are also indications (Zoccali, private communication)
that the MD of the bulge stars, when achieved with high-resolution
spectroscopy, is comparable to or narrower than that showed here, and
that there are a very few supersolar-metallicity stars.

In the same figure we also show the data from FMR06a,
who performed a new high-resolution ($R\sim 45000-60000$) analysis of
27 K-giants in the Baade's Window ($b=-4^{\circ}$) sample with the
HIRES spectrograph on the Keck I telescope and, after determining
their Fe abundances, they used them as reference stars in order to
re-calibrate the K-giant data from Rich (1988).
They found that the derived MDs are slightly stretched toward both the
metal-poor and metal-rich tails with respect to those derived in
previous works (Rich, 1988; MR94; Sadler et al., 1996; Zoccali et al.,
2003), although the overall consistency among these different MDs is
reasonable. The metal-rich end probably extends to spuriously high
abundances.

\subsubsection*{$\alpha$-element and carbon abundances}

\begin{itemize}
\item[-] Abundances of O, Mg, Si, Ca, C and Fe for stars and clusters
  in the bulge are taken from Origlia et al. (2002), Origlia \& Rich
  (2004), Origlia et al. (2005) and Rich \& Origlia (2005). We refer
  to these abundances as the ``IR spectroscopic database'' hereafter.         
  These datasets were obtained using the NIRSPEC spectrograph at Keck
  II, which allowed the determination of near-infrared,
  high-resolution ($R \sim 25000$), high signal-to-noise ratio (S/N $>
  40$) \'echelle spectra and were the only ones to provide C
  abundances for bulge giants. They used the $1.6 \mu$m region of
  the spectrum, corresponding to the H band.
  In all cases, abundance analysis was performed by means of full
  spectrum synthesis and equivalent width measurements of
  representative lines.
  Reliable oxygen abundances were derived from a number of OH
  lines; 
%rather than from single atomic lines 
  similarly, the C abundance was derived from CO molecular lines,
  whereas strong atomic lines were measured for Mg, Si, Ca, Ti and Fe. 
%  However these lines are heavily saturated for light elements and
%  do not allow a very precise abundance determination if compared to
%  molecular lines.
  The data include observations of bright giants in the cores of the
  bulge globular clusters Liller 1 and NGC 6553 (Origlia et al.,
  2002, see also Mel\`endez et al. 2003),  Terzan 4 and Terzan 5
  (Origlia \& Rich, 2004), NGC 6342 and  NGC 6528 (Origlia et al.,
  2005, see also Carretta et al. 2001; Zoccali et al. 2004)
  and measurements of abundances of M giants in Baade's window (Rich
  \& Origlia, 2005). 
  The typical errors are of $\pm 0.1$ dex.
  The main considerations that were drawn from these abundance
  analyses are that $\alpha$-enhancement is safely determined in old
  stars with [Fe/H] as high as solar, pointing toward early formation
  and rapid enrichment in both clusters
  and field, which are likely to share a common formation history. 
  The [C/Fe] abundance ratios can be depleted up to a factor of
  $\approx$3 with respect to the solar value, as expected because of
  the first dredge-up and possibly extra-mixing mechanisms due to {\it
  cool bottom processing}, which are at work during the evolution along
  the RGB, as also indicated by the very low ($<10$) $\rm
  ^{12}C/^{13}C$ abundance ratio (see also Origlia et al. 2003 and
  references therein).
  The analysis of M giants yielded abundances similar to those obtained
  with high-resolution optical spectroscopy of K giants; there is an
  apparent lack of supersolar-[Fe/H] stars, but the sample is too 
  small to draw firm conclusions.
\item For O, Mg, Si and Ca we also included the abundance measurements
  of Fulbright et al. (2006b, FMR06b), who used the same spectra as in
  FMR06a, i.e. obtained the spectra of 27 bulge K giant stars at the
  Keck I telescope using the HIRES \'echelle spectrograph with high
  resolution ($R\sim 45000-67000$) and high signal-to-noise ratio. The
  typical errorbars are of  $\sim0.1$ dex.
  The outcome of their analysis is that all elements produced from
  massive stars (i.e. $\alpha$-elements, plus Na and Al) show
  enhancement in bulge stars relative to both Galctic thick and thin
  disk, although oxygen shows a sharply decreasing trend for
  supersolar [Fe/H], which is
  attributed to a metallicity-dependent modulation of the oxygen yield
  from massive stars. 
  These results suggest that massive stars contributed more to
  the chemical enrichment of the bulge than to the disk, and
  consequently that the timescale for bulge formation were shorter
  than that of the disk, although they did not exclude other
  possibilities (such as, e.g., an IMF skewed to high masses).
\item Finally, oxygen data from Zoccali et al. (2006, Z06) were also
  taken. In this paper, Fe and O abundances for 50 K giants in four
  fields ($b=-6$; Baade's Window; Blanco $b=-12$; NGC 6553) towards
  the Galactic bulge were derived; oxygen abundance was measured from
  the forbidden line at $6300\AA$. A high resolution ($R=45000$) was
  achieved with FLAMES-UVES at the VLT. The typical errorbars are of
  $\sim0.1$ dex.
  Also in this case, [O/Fe] is found to be higher in bulge stars than
  both in thick and thin disk, and supports a scenario where the bulge
  formed before the disk and more rapidly, with a formation
  history similar to that of old early-type galaxies.
  
\end{itemize}

\subsubsection*{N vs. O}

Our nitrogen abundances in the bulge are derived from PNe. 
This may represent a problem, since while the measured oxygen
abundance represents the true value of the ISM out of which the PN
progenitor was formed, the observed N abundance has contributions both
from the pre-existent nitrogen and from that produced by the star
itself during its lifetime and dredged-up to the stellar surface.
Moreover, these data come from emission lines which have a very
complicated dependence upon several parameters (such as temperature,
density and metal content) and assumptions (e.g. the photoionization
model, from where the largest uncertainties come). 
Therefore it might be dangerous to employ these measurements to trace
galactic chemical evolution.
In principle, it would be possible to discriminate between enriched
and primordial N by means of the C/N ratio (which is very different in
the two cases, being much lower in the case of nitrogen-enriched
stars). 
Unfortunately, measurements for carbon in bulge PNe are available to
date only for a very limited set of objects (Webster, 1984; Walton et
al., 1993; Liu et al., 2001). Another possibility is to make use
  of symbiotic stars, i.e. presumably M giants in binary systems with
  a white dwarf or another hot companion. The envelope of the
  symbiotic star is photoionized from the hard UV radiation, leading
  to recombination line emission. Since the envelope has being
  observed, the abundances may be less evolved than 
  in the PNe (Nussbaumer et al., 1988). 

It is possible that the N enrichment is not very dramatic
especially in non-Type I PNe, which constitute about 80\% of the
PN population in the Galaxy (Peimbert \& Serrano, 1980). 
Moreover, since the bulge presumably has not formed stars for a long
time, Type I PNe (which have high-mass progenitors and are the
most nitrogen-enriched) are not expected to be frequent. 
This is also confirmed by Cuisinier et al. (2000) who found quite
  low [N/O] ratios in their sample if compared to that resulting from
  self-enrichment. Moreover, Luna \& Costa (2005) measured the N/O
  ratio of 43 symbiotic stars towards the Galactic bulge and, 
  as can be seen in their figure 5, the values of log(N/O) are
  consistent with those coming from studies of PNe. 

We employed the compilation of G\'orny et al. (2004), who observed 44
PNe in the direction of the Galactic bulge with the aim of discovering
Wolf-Rayet stars at their centre.
The spectra were obtained with the 1.9-m telescope at the South
African Astronomical Observatory, with an average resolution of 1000.
Furthermore, G\'orny et al. (2004) also merged their data with other
published ones. Namely, they included the samples observed by
Cuisinier et al. (2000), Escudero \& Costa (2001) and Escudero et 
al. (2004). 
They obtained a total of 164 objects, among which a clear segregation
of the subsamples is seen, due to the different selection criteria
adopted to define each sample, and therefore none of them is truly
representative of the bulge PN population. 
By merging the datasets, a more complete view of this population is
achieved.
Updated reddening corrections were applied to the objects from
Escudero \& Costa (2001) and Escudero et al. (2004). 
The merged sample was divided into two classes (according to the
criteria listed in Stasi\'nska et al, 1991), the first including those
objects which are likely to be physically related to the Galactic
bulge and the second containing the remaining objects which most
probably belong to the disk.
We only selected objects belonging to the bulge which had a clear
detection of oxygen and nitrogen emission features; the resulting
sample includes 103 objects.
Errors in abundance derivations from both observational and
theoretical uncertainties are typically $0.2-0.3$ dex for [O/H] and
can be even larger for [N/O]. 

In the future, when very high resolution IR spectroscopy ($R
\approx$100,000) becomes possible, N estimates could be also derived
in giant stars from the faint CN lines that are de-blended from
stronger the CO and OH lines at high resolution.

\section{Model results}

\subsection{Supernova rates}
\label{sec:snrate}

\begin{figure*}
\centering
\includegraphics[width=.5\textwidth]{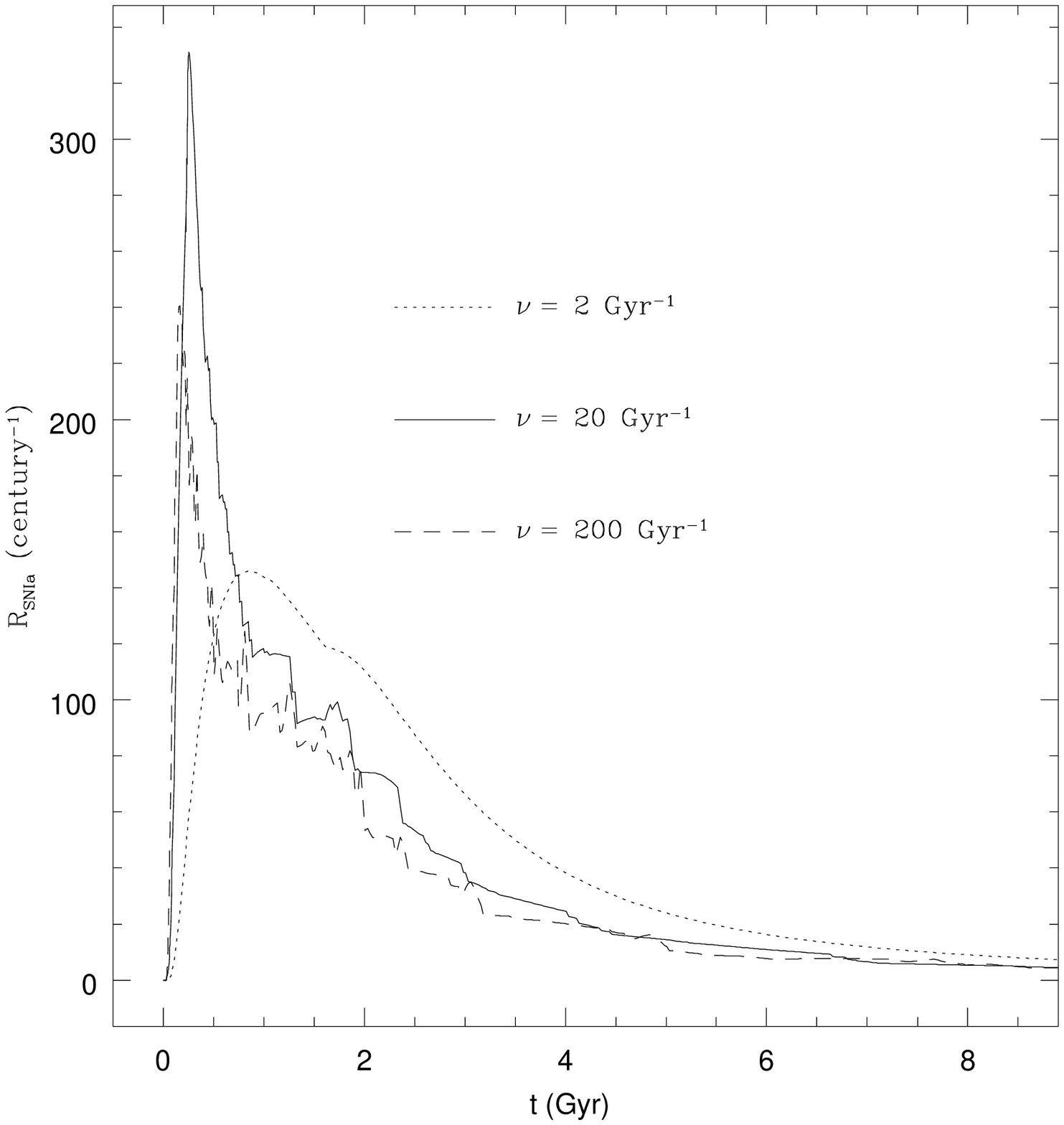}%
\includegraphics[width=.5\textwidth]{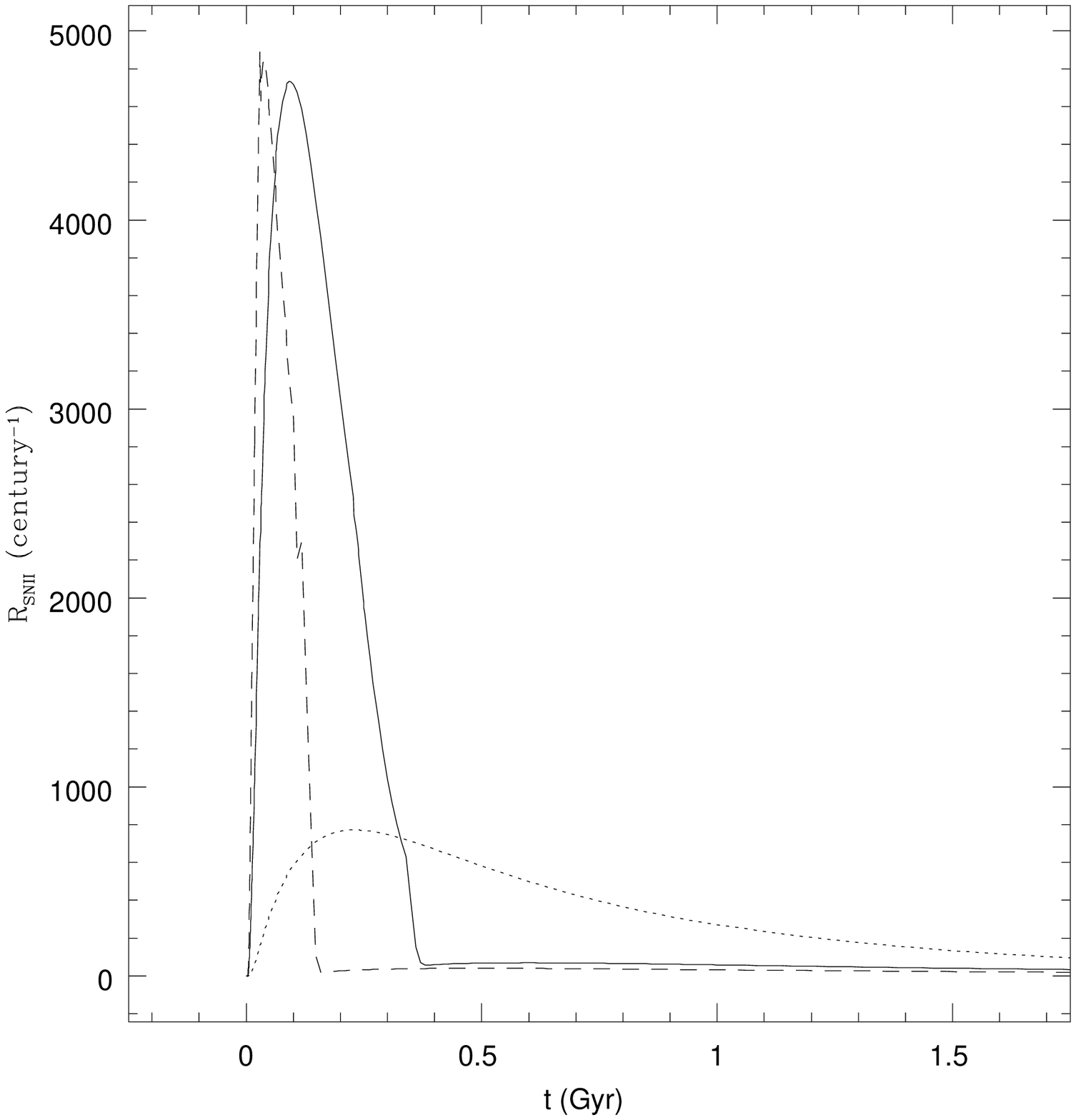}
\caption{Evolution of the rates of Type Ia (left) and Type II (right)
  supernovae where different values of the star formation efficiency
  (middle panel).}
\label{fig:fig02}
\end{figure*}

Fig. \ref{fig:fig02} shows the predicted time evolution of the rate of
Type II and Type Ia SNe in the Galactic bulge; the former, that die on
small timescales, closely reflect the evolution of the SFR (for
simplicity, we only show the case of different star formation
efficiencies). 
The secondary peaks of the SNIa rates are
software-related and are mostly due to the 
discontinuities in the adopted stellar lifetimes (Kodama, 1997) but
do not affect the results concerning chemical abundances.
The break in the Type II SN rates corresponds to the suppression of
the star formation rate due to the achievent of the condition
expressed in Eq. \ref{eq1}, which quite intuitively occurs earlier for
flatter IMFs, higher $\nu$'s and/or lower $\tau$'s.
However, even without a galactic wind, the Type II SN  rate would
become negligible at the same epoch, owing to the small amount of gas
left in the bulge at that time. 

In those cases where the star formation is ``bursty''
(e.g. high star formation efficiency and quick formation timescale),
the peak of the Type Ia SN rate (and therefore of Fe enrichment) can occur 
even before 1 Gyr, which is the timescale for Fe-enrichment in the
solar neighbourhood. 
In fact, the time of occurrence of this peak is very sensitive to the
underlying star formation history and can be as low as $\sim 0.2$ Gyr
in the case of $\tau=0.01$ Gyr or $\nu=200$ Gyr$^{-1}$ (see also
Matteucci \& Recchi, 2001).   
Therefore, we expect that: 
\begin{enumerate}

\item The down-turn (i.e. change of slope) of the [$\alpha$/Fe] vs. [Fe/H]
plots,
  corresponding to the beginning of a sustantial Fe enrichment, 
 will occur later, in
  general, at higher [Fe/H] values relative to the change in slope in
  the solar    vicinity. This is due to the high star formation rate
  and the short timescale for bulge formation. This fact was 
  pointed out for the first time by MB90.

\item Where the IMF is flatter, the abundance trend of
  $\alpha$-elements vs. Fe is flatter as well, because SNeIa 
are fewer than in a Salpeter or steeper IMF.
 We also expect that in the Z00-4 case,
  which favours the creation of SNIa progenitors, Fe production will
  be enhanced. 

\item For steeper IMFs (e.g. Scalo, 1986, or Salpeter, 1955), the
  nitrogen enrichment from low- and intermediate-mass stars is
  enhanced with respect to O, which mostly comes from massive
  stars. 
  As a result, the [N/O] vs. [O/H] plot should lie above the one of
  our fiducial model. 
  Moreover, N primary production from low- and intermediate- mass
  stars should be enhanced as well, and the flattening of the above
  mentioned plot towards the latest evolutionary stages should be
  prolonged. 
  In general, any model that favours the contribution of massive
  rather than low- and intermediate- mass stars (i.e. high $\nu$, low
  $\tau$) should predict a downturn of  the [N/O] vs. [O/H] plot if
  compared to our fiducial model, and vice versa. 
\end{enumerate}

\subsection{Metallicity distribution of bulge giants}
\label{subs:GKM}

\begin{figure}
\centering
\includegraphics[height=.3\textheight]{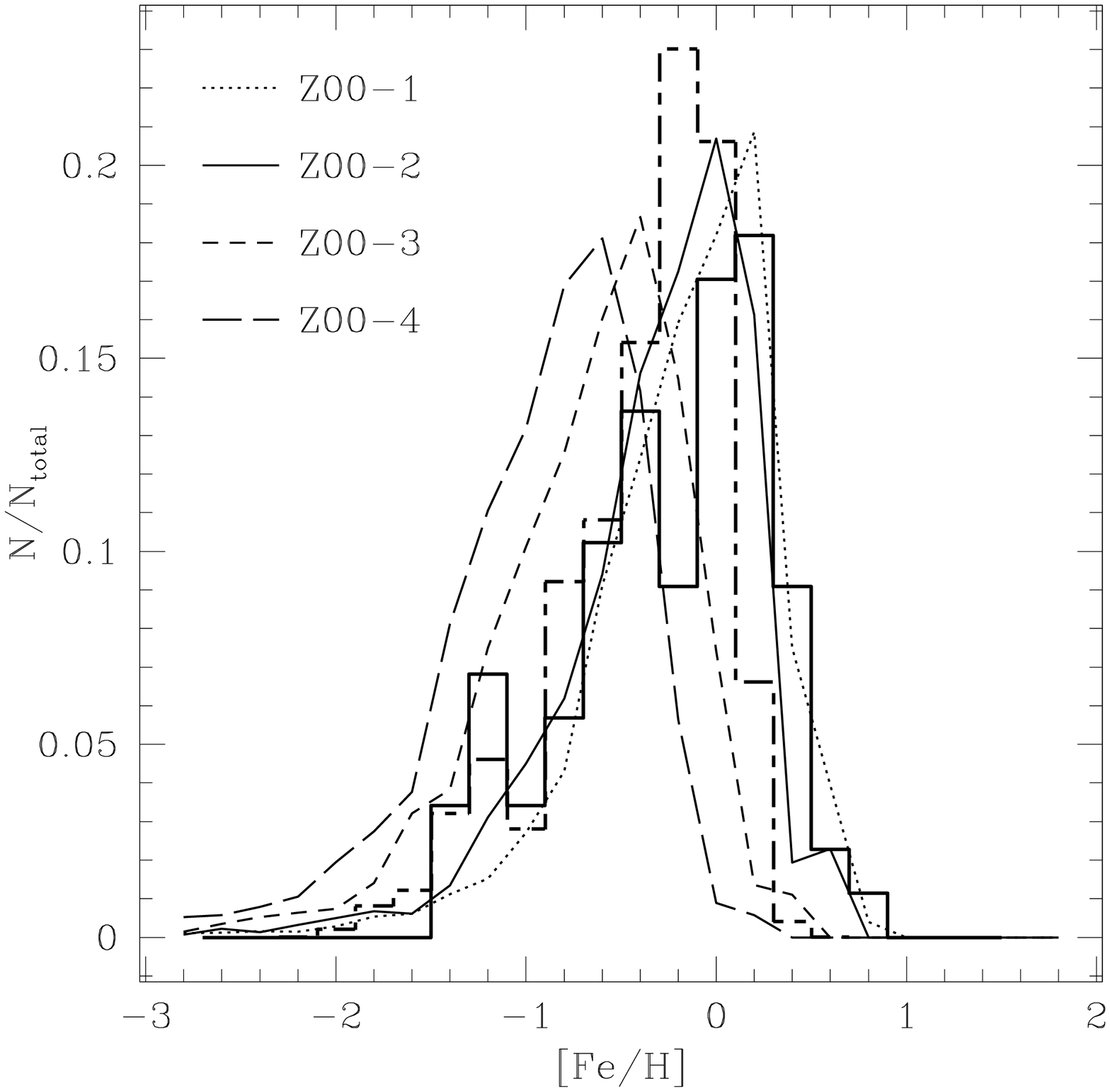}\\
\includegraphics[height=.3\textheight]{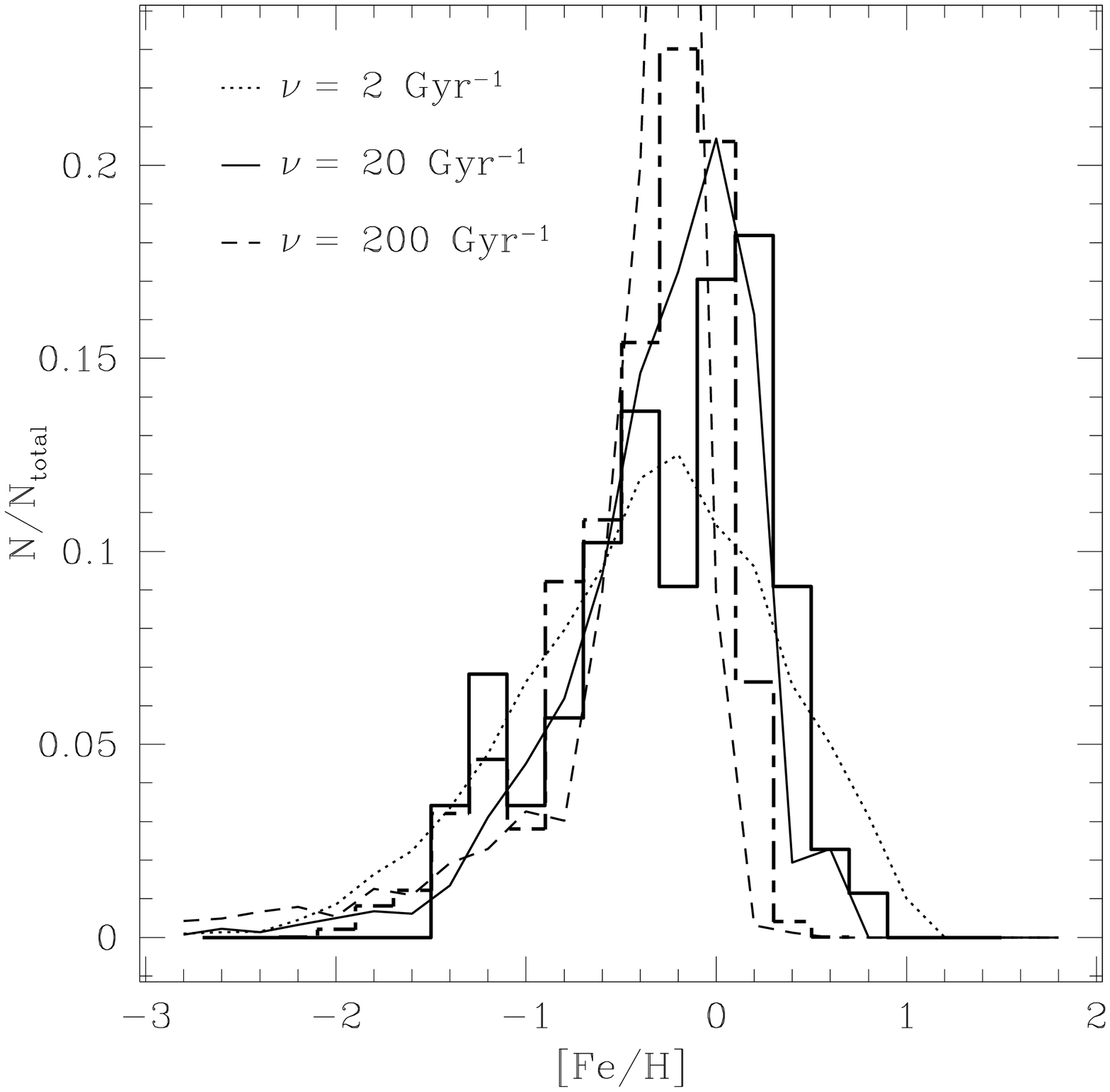}\\
\includegraphics[height=.3\textheight]{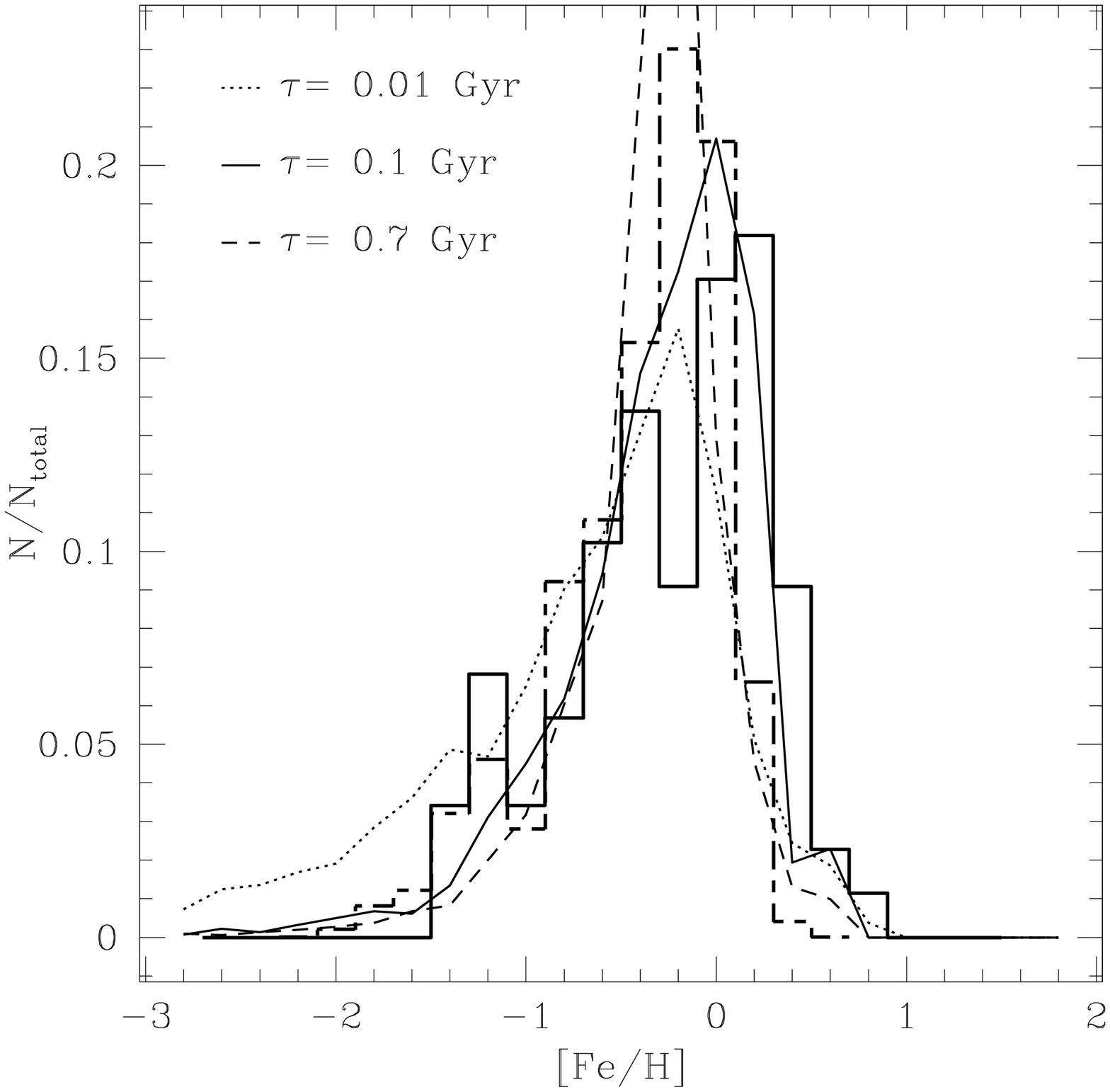}
\caption{Predicted metallicity distributions in our bulge models for
  single parameter variations relative to the reference model, with
  different choices of the IMF index (upper panel; see text for details),
  star formation efficiency (middle panel), and gas infall timescale 
  (lower panel). Our fiducial model is represented by the solid
  line.  Data
  are from Z03 (solid histogram) and FMR06a (dashed histogram).}  
\label{fig:fig03}
\end{figure}

Fig. \ref{fig:fig03} shows the predicted MD for giant stars
compared to the data from Z03 and FMR06a\footnote{We must remark
that we did not make an attempt to convolve the predictions with
  uncertainties. Since the observed distributions are probably broader
  than the ``true'' distribution, due to random errors, and may be
  offset due to systematic errors, their comparison is likely to be
  affected.}. 

The choice of a different IMF does not primarily influence the spread
of the distribution but rather shifts its peak on the [Fe/H] scale:
in general, flatter IMFs result in distributions peaked at 
higher values of [Fe/H].
Since the star-forming phase is relatively short, massive stars will
be important in the Fe enrichment; if we adopt the Scalo (1986)
exponent (model Z00-4) rather than the Salpeter (1955) one
(model Z00-3) for stars more massive than $2 M_{\odot}$, the
resulting MD will change dramatically, moving the peak toward
significantly lower [Fe/H].
%To check this point, we also tested a model where another two-slope
%IMF was adopted, with $x=0.95$ instead of $x=1.35$ for $1 M_{\odot} <
%M < 2M_{\odot}$; the resulting distribution, which we did not show to
%avoid overcrowding the figure, is very similar to that obtained with
%the Scalo IMF. 
Considerations on the MD allow us to exclude IMFs steeper than
$x=0.95$, whereas plots for flatter IMFs are all consistent with the
observed distributions.
In fact, the MD obtained with the very flat ($x = 0.33$) IMF of case
Z00-1 is almost undistinguishable from that calculated in case Z00-2,
showing that the MD progressively becomes almost insensitive to 
the flattening of the IMF under a certain value of the index.
This is maybe suggestive of a ``saturation effect'' in the Fe
enrichment from Type II SNe.
Namely, a flattening of the IMF below a certain value of the
  index $x$ does not produce a sensible increase of the number of
  massive stars, and moreover, the Fe mass ejected is assumed to be
  the same (i.e. does not increase as a function of mass) for all
  masses above $40 M_{\odot}$ (see \S \ref{sec:model}).

If we choose a moderate star formation efficiency
($\nu=2$ Gyr$^{-1}$, comparable to that of the galactic halo), the predicted
metallicity distribution is broadened and 
the number of both high- and low-metallicity stars is somewhat
overestimated.  
The observed peak is poorly reproduced (both in height and position)
for all of the MDs considered.
In contrast, the extremely high value of $\nu=200$ Gyr$^{-1}$
yields an extremely narrow distribution, with an excessive peak height
(reaching 0.68, but is truncated in the figure to preserve clarity). 

A change in the infall timescale affects mainly the spread of
the distribution and only slightly the position of the peak. 
The low-metallicity wing is especially sensitive to the value of
$\tau$.
The model with $\tau=0.01$ Gyr approaches a closed-box model, i.e. a
model where all the gas is already present from the beginning (which 
would correspond to the limit $\tau=0$). 
The gas soon reaches high densities and is consumed very rapidly. 
Thus, the number of metal-poor stars is overestimated and the
predicted distribution extends below the observed low-metallicity
tail. 
This confirms the considerations of Z03, i.e. an equivalent (although
less important) manifestation of the G-dwarf problem occurs also in
the Galactic bulge.
We also calculated the MD resulting from the adoption of a slightly
longer infall timescale, namely $\tau=0.7$ Gyr, that corresponds to
the timescale $\tau_H$ for collapse from halo to bulge in the model of 
Moll\'a et al. (2000). 
As the figure shows, the $\tau=0.7$ Gyr MD is at variance with
observations, having a serious deficit of both metal-poor and
metal-rich stars and too
high a peak with respect to both measured distributions (again higher
than 0.5 but truncated in the figure).

The main conclusions which can be drawn for variations of a single
parameter are:

\begin{enumerate}
\item Changing the IMF slope has the general effect of shifting the
  peak of the MD towards lower metallicities for steeper IMFs, and
  vice versa. For $x \leq 0.95$ such variations become much less
  evident.
  We want to point out that continuous wind such as that of the
  Samland et al. (1997) model cannot have the same effects of
  flattening the IMF, since outflows \emph{lower} the effective yield,
  and thus if we had a continuous outflow we would need an even flatter
  IMF to reproduce the observed MD and abundance ratios. This was
  shown i.e. for the galactic disk by Tosi et al. (1999, their figure 7).
\item Lower star formation efficiencies give rise to broader MDs and
  viceversa. The same is true of shorter infall timescales, which
  additionally increase the number of low-metallicity stars.
\item The effect of both star formation efficiency and infall
  timescale on the position of the MD peak is negliglible.
\end{enumerate}

We now explore whether any other combinations of parameters can
fit the observed MD. Since the position of the peak is
essentially only determined by the IMF, combining IMFs other than
those of models Z00-1 and Z00-2 with different values of the other
parameters would not be useful to reproduce the required MD.
On the contrary, the effects of infall timescale and star formation
efficiency could compensate each other; therefore, we also considered
the following ``supplementary'' (S) models:

\begin{itemize}
\item[-] \emph{Model S1}: IMF like in model Z00-1, $\nu=200$ Gyr$^{-1}$,
  $\tau=0.01$ Gyr.
\item[-] \emph{Model S2}: same as in S1, but with IMF like in model Z00-2.
\item[-] \emph{Model S3}: IMF like in model Z00-1, $\nu=2$ Gyr$^{-1}$,
  $\tau=0.7$ Gyr.
\item[-] \emph{Model S4}: same as in S3, but with IMF like in model Z00-2.
\end{itemize}

\begin{figure*}
\centering
\includegraphics[width=.5\textwidth]{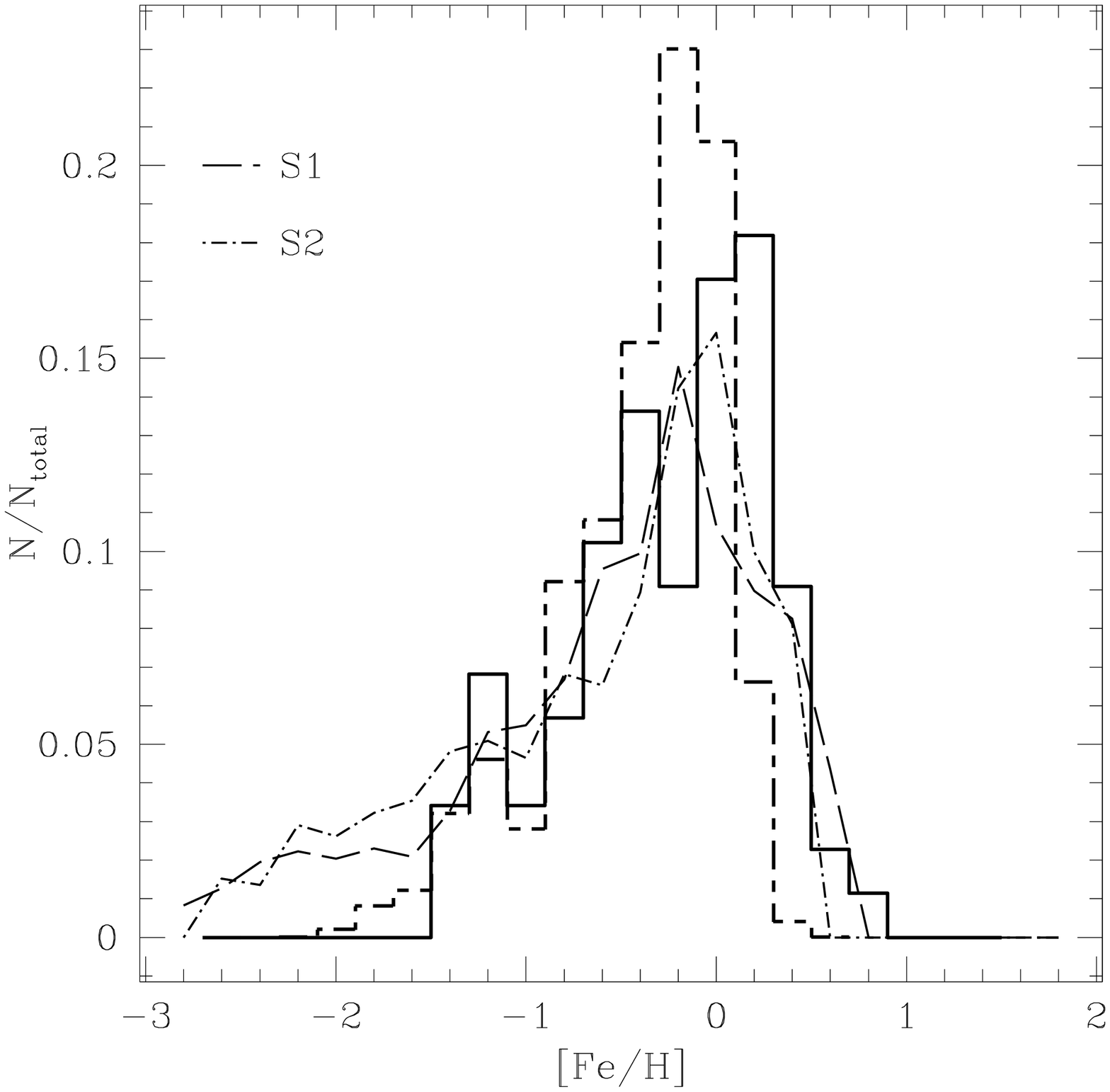}%
\includegraphics[width=.5\textwidth]{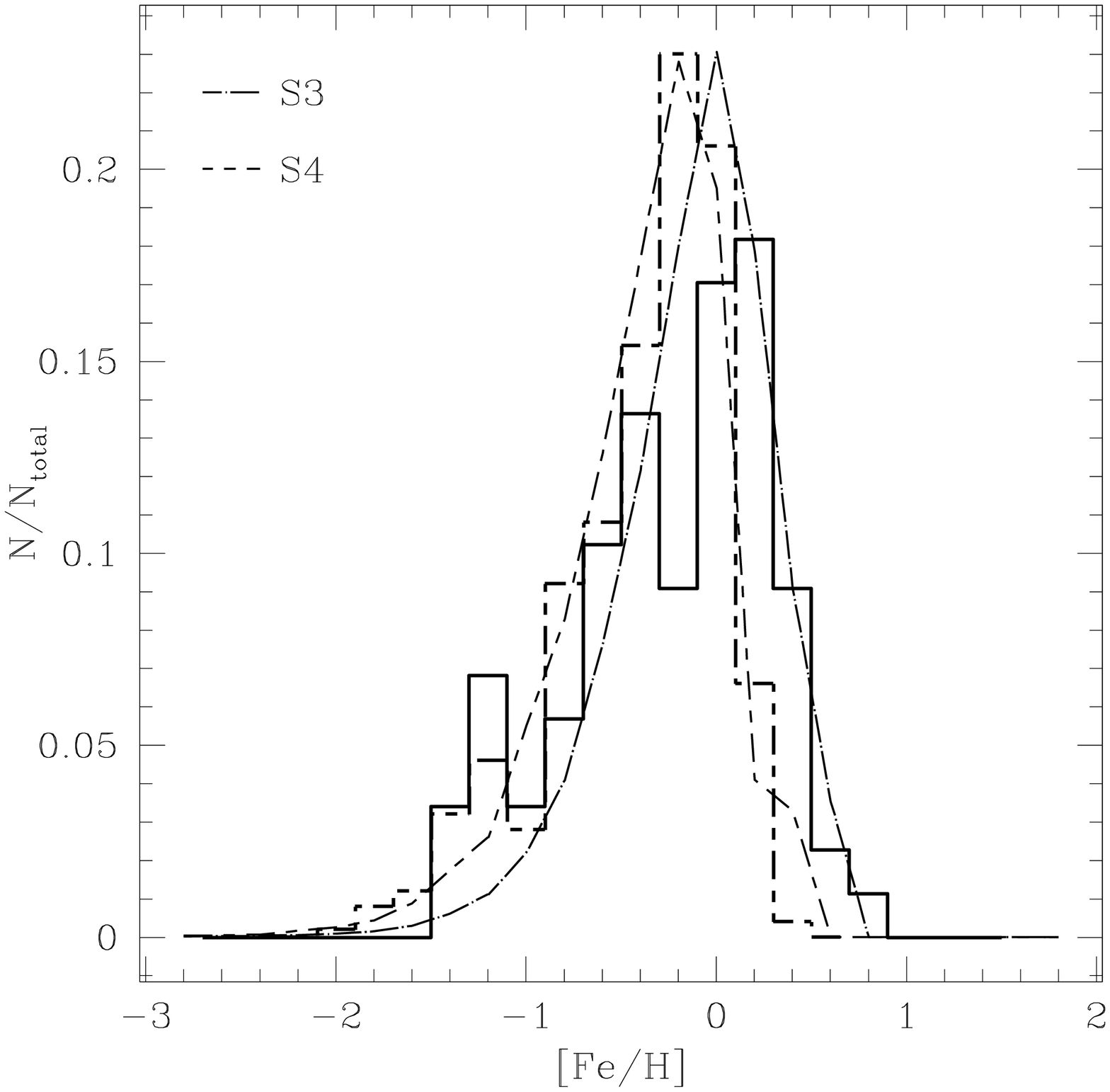}
\caption{Predicted MDs in models with $\nu=200$ Gyr$^{-1}$ and
  $\tau=0.01$ Gyr (left panel) and in models with $\nu=2$ Gyr$^{-1}$
  and $\tau=0.7$ Gyr, with the adoption of two different IMFs (see
  text for details). Data are from Z03 (solid histogram) and FMR06a 
  (dashed histogram).}
\label{fig:fig04}
\end{figure*}

In Fig. \ref{fig:fig04} are shown the resulting MDs for these
supplementary models.

It is clear that while combining a high value of $\nu$ and a low
value of $\tau$ exacerbates the bulge ``G-dwarf problem'', a 
 longer formation timescale can combine with a milder star formation
 efficiency to give rise to a MD compatible with observations, and
 this is what happened in some of the previous models (Samland et al.,
 1997; Moll\'a et al., 2000).
However, such a degeneracy cannot be pushed above a certain range of
values. 
We investigated longer timescales and found out that it becomes
progressively harder to mantain the required MD both in height and
position, since the result of combining much longer timescales with
efficiencies suitable to keep the required shape is to shift the
distribution with the required shape toward lower metallicities,
and vice versa, if we try to keep the predicted MD at the right
position, its shape becomes too narrow, somewhat in contrast with the
observed MDs.
In any case, as we shall see, such a degeneracy is definitively broken
when  we take into account the evolution of abundance ratios (see next 
  Section).

For $\tau \geq 2$ Gyr, some star formation is predicted at detectable
levels at the present time, contrarily to observations. 

%Moreover, the value of $\nu$ necessary to fit the MD with $\tau
%\gtrsim 1$ Gyr is less than $1$ Gyr$^{-1}$, i.e. lower than for the
%solar neighbourhood, which seems rather implausible.

Therefore, we conclude that good agreement with the observed MD is
achieved only if the bulge formed on a short timescale, with a flat IMF
and with a rather high star formation efficiency.

\subsection{Evolution of abundance ratios}
\label{sec:abr}

\subsubsection*{Oxygen}

\begin{figure*}
\centering
\includegraphics[width=.95\textwidth]{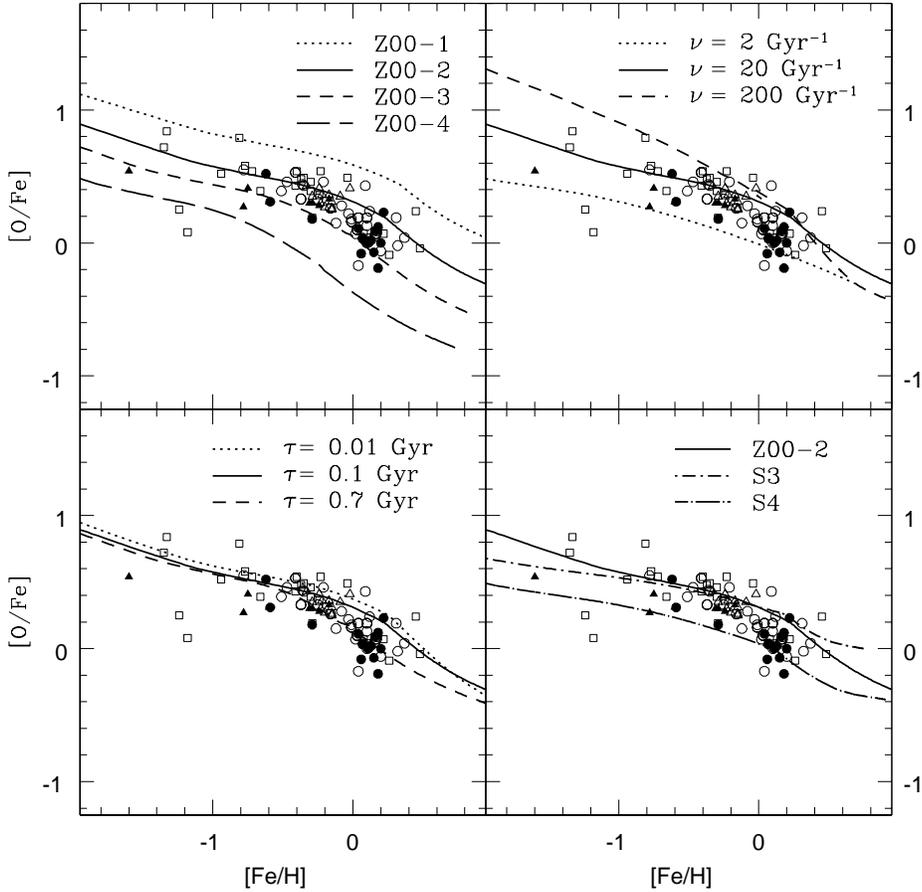}
\caption{Evolution of [O/Fe] vs. [Fe/H] in the bulge for different
  values of the IMF index (upper left panel; see text for details),
  star formation efficiency (upper right panel), and infall timescale
  (lower left panel). 
  The lower right panel shows the results obtained
  combining a longer infall timescales and milder star formation
  efficiencies for two different IMFs (see text for details). 
  Triangles represent data taken from the IR spectroscopic
  database (field stars are symbolized by filled triangles, stars in
  globular clusters by open triangles. 
  See \S \ref{sec:data} for specific references).
  The open and filled circles represent the low- and high-quality
  data, respectively, from Zoccali et al. (2006). Finally, the
  squares are the data from Fulbright et al. (2006b) and Zoccali et
  al. (2006).}
\label{fig:fig05}
\end{figure*}

Figure \ref{fig:fig05} shows the evolutionary plots of the
[O/Fe] abundance ratio with metallicity, which is the most sensitive
to the adopted parameters. 
The models are compared to the dara of the IR spectroscopic database,
FMR06b and Z06.
Each panel illustrates the effects of
changing one of the model parameters as stated in \S\ \ref{sec:model}. 
The last panel also shows the results obtained with models S3 and S4.
We remind that when the bulk of SNeIa begin to
explode, the [O/Fe] ratio has a downturn and a knee in the
[O/Fe] vs. [Fe/H] curve is expected.

The fiducial model provides a satisfactory fit to the existing
measurements and predicts the correct amount of oxygen enhancement,
even though the slope of the predicted plot in the fiducial model
  is slightly flatter than the observations for supersolar [Fe/H].  
  However, as we shall see, the fit to this abundance plot cannot be
  improved without violating other constraints. 
The amount of $\alpha$-enhancement crucially depends on the IMF index,
and more in particular on the IMF of the Type II SN progenitors.
The larger the number of massive stars, the higher the plateau at low
metallicities.
Model Z00-1 overproduces oxygen, and is not consistent with data.
Fig. \ref{fig:fig05} also allows us to exclude a Scalo (1986)
exponent (model Z00-4) for massive stars in the bulge, since the
corresponding evolution of [$\alpha$/Fe] vs. [Fe/H] lies well below
the observed data points. 
A Salpeter (1955) IMF (model Z00-3) cannot be excluded on the
basis of these abundance ratios(; however, it was ruled out by the
  MD plot. 

Concerning the star formation efficiency, a value of $\nu = 20$
Gyr$^{-1}$ is consistent with the observed abundance ratios, whereas
lower or higher values seem to be at variance with the (few)
lowest-metallicity data. 
The model with $\nu = 200$ Gyr$^{-1}$ predicts values of [O/Fe] which
are larger than those of the fiducial model at low metallicities, but
then the gas is consumed very rapidly and star formation cease;
this gives rise to a sharper decrease in the [O/Fe] ratio at higher
metallicities. The predicted slope is compatible with
  observations, however the absolute amount of oxygen enhancement is
  slightly overestimated, and in any case this model was already
  excluded on the basis of the MD (Fig. \ref{fig:fig03}).
It is noteworthy that even though the star formation stops  much
earlier in time for higher $\nu$'s, the model trajectory still spans
the same range in [Fe/H] for all values of the star formation
efficiency $\nu$. This is because at high values of the star formation
efficiency, higher metallicities are attained in shorter times.

On the basis of the MD, we already saw that models with
$\nu=2$ and 200 Gyr$^{-1}$ must be excluded; considerations derived
using the [O/Fe] abundance ratio provide a useful consistency check. 
We then suggest that a value of $20$ Gyr$^{-1}$ can fit both the MD
and the abundance ratios.

Instead, changing the infall timescale from 0.01 to 0.7 Gyr has
almost no effect on the evolution of the [O/Fe] abundance ratio, and
this holds also for other $\alpha$-elements, if we exclude a
 small improvement of the agreement with data in the $\tau = 0.7$ Gyr
 case (which however, as well as the $\nu = 200$ Gyr$^{-1}$ case, is
 excluded by the MD plot).
Therefore, there is no way of distinguishing between these models on
the basis of considerations about abundance ratios, although we
already excluded the cases of $\tau \ll 0.1$ Gyr and $\tau > 0.1$ Gyr
which yielded MDs in contrast with the observed ones.

Finally, we discuss the results obtained with the ``supplementary''
models S3 and S4, whose MD was consistent with observations. We can
see that it is necessary to assume a very flat IMF (model S3) to
  avoid underestimating the [O/Fe] ratio with the adoption of
  $\tau=0.7$ and $\nu=2$ Gyr$^{-1}$, and the predicted slope is
    much flatter than the observed one. For even longer timescales,
  flattening the IMF is no longer sufficient to fit the data. This 
  is due to the fact that Type Ia SNe have more time to pollute the
  ISM with Fe causing the plot to turn down, falling below the
  observations. Therefore, formation timescales longer than $\sim1$
  Gyr must be excluded in any case. This is the consequence of the
  time-delay model for Type Ia SNe in the case of different star
  formation histories, as already pointed out by MB90 and Matteucci
  (2000).

\subsubsection*{Other $\alpha$-elements}

\begin{figure*}
\centering
\includegraphics[width=.95\textwidth]{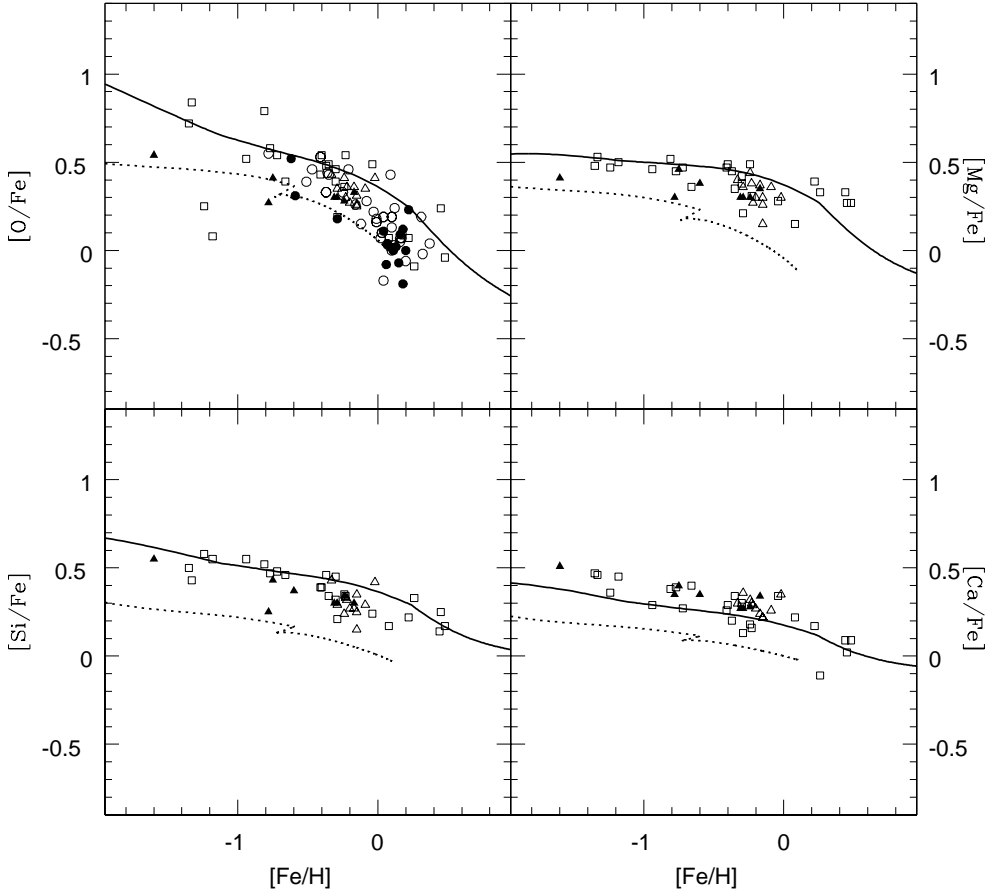}
\caption{Evolution of [$\alpha$/Fe] vs. [Fe/H] in the bulge for O, Mg,
  Si, Ca for the fiducial model (solid line). We also plotted a solar
  neighbourhood fiducial line (dotted line) for comparison. Data for
  oxygen are the same as in Fig. \ref{fig:fig05}. For the other
  elements, data are taken from Fulbright et al. (2006b, open
  squares) and, for Mg, from the IR spectroscopic database (field
  stars: filled triangles, clusters: open triangles. See text for
  detailed references).}
\label{fig:fig06}
\end{figure*}

In figure \ref{fig:fig06} are shown the evolutionary plots of the
abundance ratios of several $\alpha$-elements (O, Mg, Si and Ca) to
Fe versus Fe abundance, compared to the same datasets as [O/Fe], with
the exception of Z06. 
We also show a solar neighbourhood fiducial line for
  comparison, calculated according to Fran\c cois et al. (2004). 
A striking aspect of the predicted [$\alpha$/Fe]
vs. [Fe/H] relation is that the slope of the [$\alpha$/Fe] ratios changes
only at [Fe/H] $\simeq 0$ for the fiducial model, in agreement with
the data and at variance with the solar vicinity where this occurs at
[Fe/H] $= -1.0$.
Indeed, it is evident that a star formation history and an IMF such as
those suitable for the solar neighbourhood give results which do not
agree with the bulge observations, the general trend being to severely
underestimate the data especially at [Fe/H]$\gtrsim -1$, when Type I
SNe, whose formation is favored by the steeper IMF, start polluting
the ISM considerably.

%The fiducial model predicts that while O and Mg are enhanced over
%the whole range of [Fe/H], Si and Ca exhibit a quite flatter trend
%which decreases more slowly with respect to the former
%$\alpha$-elements, {\bf even though this difference is not as
%  remarkable as in the disk}. Si and Ca follow more closely the
%  evolution of Fe because these elements are partly produced also in
%  Type Ia SNe,{\bf which have a larger weight in the solar
%    neighbourhood with respect to the bulge.}

  It is worth noting that our fiducial model predicts that the
  slopes of [O/Fe] and [Mg/Fe] vs. [Fe/H] are different, the latter
  being less steep. 
  This is mainly due to the fact that whereas magnesium is mainly
  produced by a restricted range of stellar masses (around $20-25
  M_{\odot}$) oxygen is produced by a broad interval of stellar masses,
  (from $10$ to $100M_{\odot}$) the O/Fe production ratio increasing
  with stellar mass.  
  This different trend is present also in the solar vicinity data
  (see Fran\c cois et al., 2004) and is confirmed in the bulge
  by the observations of FMR06, which cover a larger range of
  metallicities than the IR spectroscopica database (i.e. [Fe/H]
  $\gtrsim 0$ and [Fe/H] $\lesssim -1$).

%For flat IMFs this change of slope is less easily perceivable than for
%steeper ones, because of the minor fraction of low-mass stars giving
%rise to SNIa explosions. 
%Observations near such metallicities would help fix the point of
%change of slope and provide an additional constraint.

\subsubsection*{Carbon and nitrogen}

\begin{figure*}
\centering
\includegraphics[width=.95\textwidth]{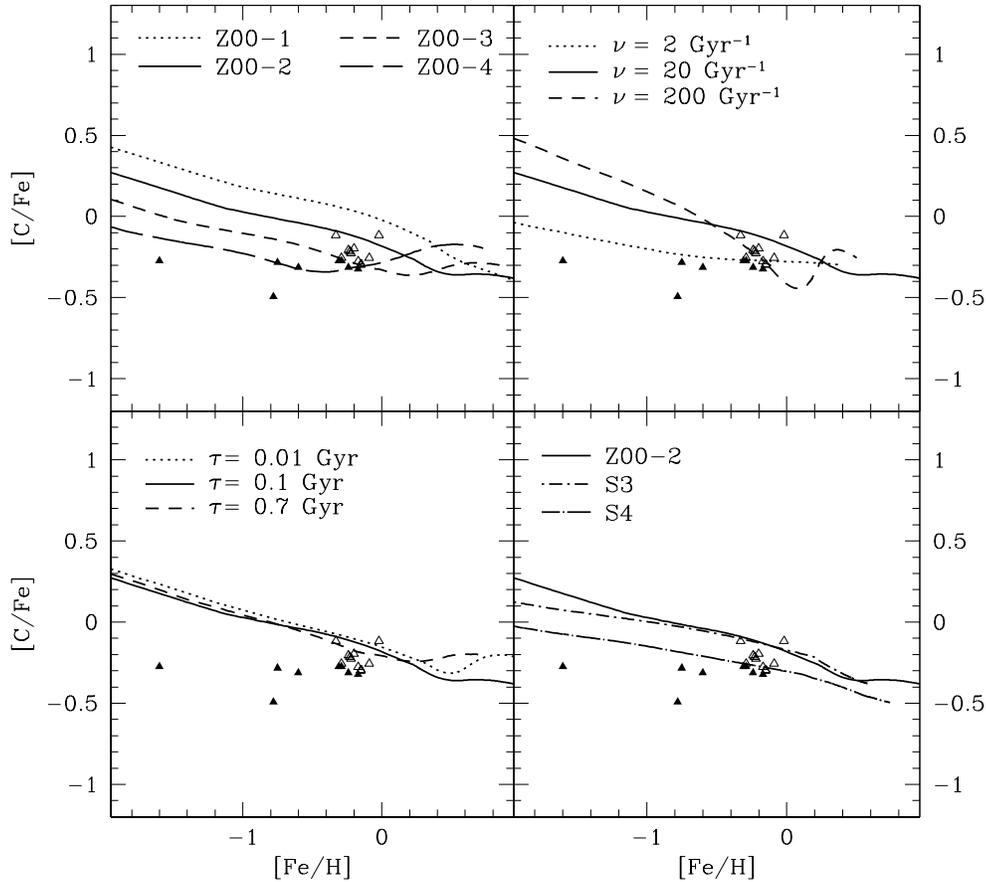}
\caption{Evolution of [C/Fe] vs. [Fe/H]  in the Galactic bulge for
  various choices of the IMF (upper left panel) star formation
  timescale (upper right panel), and infall timescale (lower left
  panel). The lower right panel shows the results obtained combining a
  longer infall timescale and a milder star formation efficiency for
  two different IMFs (see text for details). Data sources are the same
  as in Fig. \ref{fig:fig05}. Data
  are from the IR spectroscopic database (field
  stars: filled triangles, clusters: open triangles. See text for
  detailed references).}
\label{fig:fig07}
\end{figure*}

\begin{figure*}
\centering
\includegraphics[width=.95\textwidth]{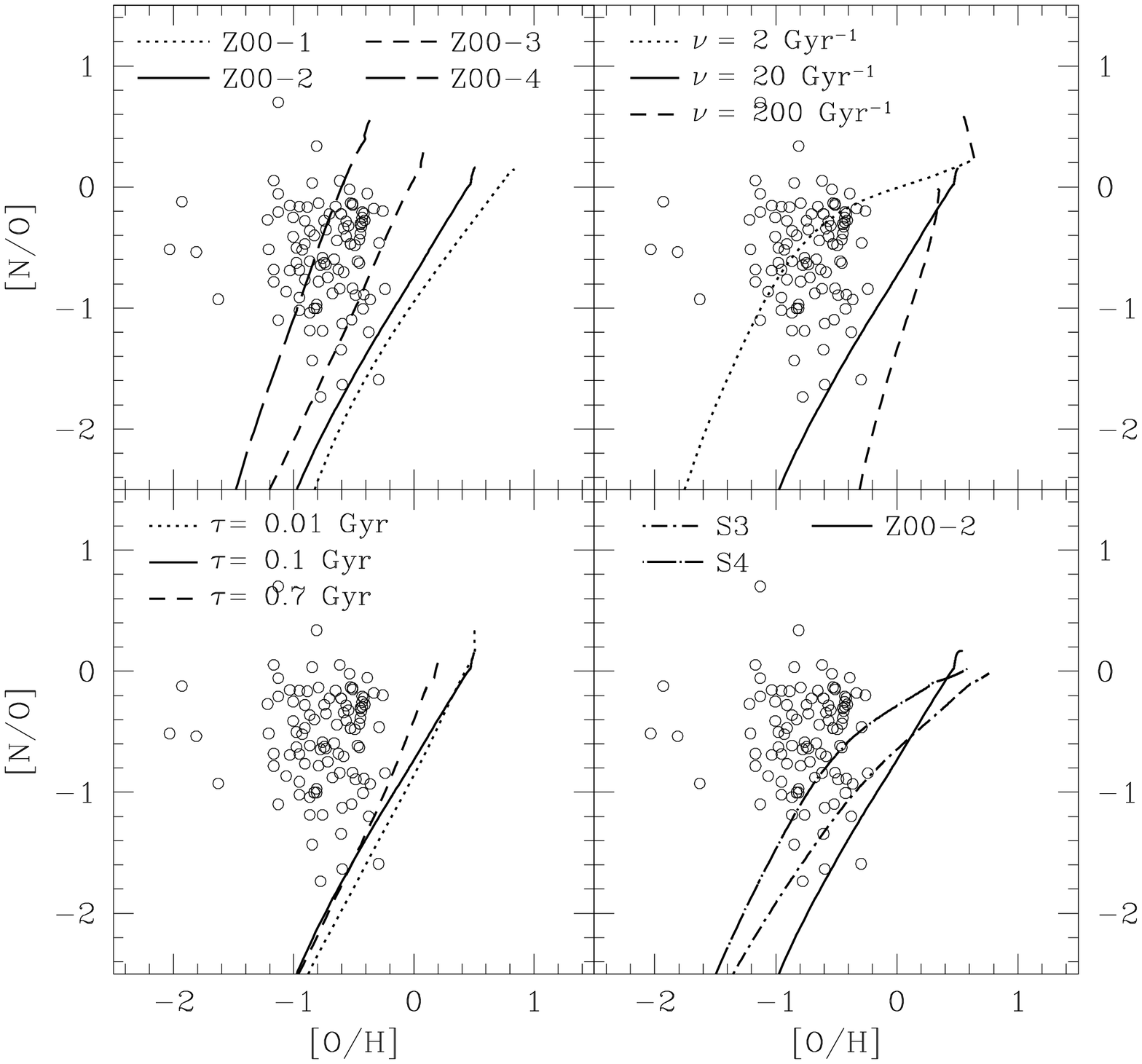}
\caption{Evolution of [N/O] vs. [O/H] in the Galactic bulge for
  various choices of the IMF (upper left panel) star formation
  timescale (upper right panel), and infall timescale (lower left
  panel). The lower right panel shows the results obtained combining a
  longer infall timescale and a milder star formation efficiency for
  two different IMFs (see text for details). The abundances measured
  from bulge PNe are taken from G\'orny et al. (2004).}
\label{fig:fig08}
\end{figure*}

Figures \ref{fig:fig07} and \ref{fig:fig08} show the
evolutionary behaviours of C and N, which, in contrast to the 
$\alpha$-elements are mainly produced by low- and intermediate-mass
stars (and therefore on long timescales).
The available dataset for carbon is limited to the measurements of 
the giant stars, which however are known to undergo mixing processes 
along the RGB, severly affecting the carbon abundances (in fact, we
can immediately see that none of the models lie beneath the
observations, since carbon is likely to be depleted during the
evolution along the RGB).
Hence, the comparison between models and observations cannot allow 
to firmly constrain the former and to reach strong conclusions in
general. The observations can only provide a lower limit to any
plausible model. 

The behaviour of [C/Fe] at low metallicities is largely affected by
the assumed star formation efficiency, just as in the case of oxygen,
whereas the curves do not show really discernible differences above
the solar metallicity. 
A high value of the star formation efficiency ($\nu = 200$ Gyr$^{-1}$)
enhances the number of stellar generations and therefore accelerates
the formation of intermediate-mass stars which are the main C
producers, but again [C/Fe] falls below the predictions of the
fiducial model due to fast gas consumption, as was the case for oxygen.
The steeply falling trend of [C/Fe] until $\sim$solar [Fe/H] is due to
the fact that the C/Fe production factor is decreasing with decreasing
mass for high-mass stars.
In this case too, abundance data for metal-poor bulge stars can 
confirm or refute the conclusions reached on the basis of the MD, even
though large values of $\nu$ seem again excluded.

A change in the IMF has a major impact on the contribution from
low-mass stars. The plots show a bump at [Fe/H] ranging from $\sim -0.4$
to 0.4, which corresponds to the time at which the first intermediate
mass stars contribute to C enrichment, with the exception of the Z00-1
IMF, for which the star formation ceases before the bulk of
intermediate-mass stars have time to die. 
This bump is not visible in the solar neighbourhood, and its 
occurrence in the bulge is related to the strong
metallicity-dependence of the adopted C yields for low- and 
intermediate-mass stars  (Van den Hoek \& Groenewegen, 1997), combined
with the fact that in the bulge high metallicities are reached very
early. 
The feature is more pronounced for larger values of $x$ because low-
and intermediate-mass star formation is favoured by these steeper
IMFs.
This is particularly evident from the difference between the plots
adopting a Salpeter (1955) and a Scalo (1986) IMF, respectively.
Before the onset of the intermediate-mass dominated regime, the amount
of C enrichment increases for flatter IMFs due to the enhanced C
production from massive stars; on the contrary, when the massive star
dominated regime ends, C production is increased remarkably for
steeper IMFs due to the enhanced contribution from low- and
intermediate-mass stars. Again, the Scalo IMF looks rather
implausible, since the predicted trend almost lies beneath the data
points, and this is likely to mean that C is underproduced in this model.

No appreciable change is seen when the adopted gas infall
timescale is varied, with the exception of a slightly different
position of the occurrence of the bump. 
The plots resulting from models S3 and S4 are also shown. Bearing in
mind the paucity of data below [Fe/H] $\sim -0.5$, the
model S4 predicts a trend which passes through the cluster of
data points at [Fe/H] $\sim 0.0$, but since as we already stated C is
likely to be affected by stellar evolution, this model probably
underpredicts the [C/Fe] ratio, again due to the Fe contribution from
Type Ia SNe. Instead, model S3 leads to a larger C production and
therefore keeps the consistency with the observations, the trend being
only flatter than that of the fiducial model.

\vspace{1pc}

Nitrogen is the only element for which our reference model cannot
yield a satisfactory agreement with data: in fact, it tends to lie
under the observational points, following their lower envelope. This
is true also for the solar vicinity.
As we can see from Fig. \ref{fig:fig08}, there is no way of
reproducing the average trend of the [N/O] abundance ratio with [O/H]
for single-parameter variations unless we adopt parameters which have
proven to lead to results at variance with other constraints: e.g. a
very low star formation efficiency, which results into an
underproduction of oxygen with respect to nitrogen and was dismissed
on the basis of the resulting MD (and also to some extent on the basis
of the [O/Fe] plot); or a steep IMF, favouring the formation of low-
and intermediate-mass stars (which are supposed to produce the bulk of
N), which did not reproduce either the MD or the evolution of
[$\alpha$/Fe] vs. [Fe/H].

A good agreement is instead achieved with the S4 case, since it
achieves the same effect of a low efficiency of star formation
(i.e. oxygen underproduction) while the longer formation timescale
does not influence the evolution of abundance ratios. 
Models S3 still somewhat improves the match with data relative to the
fiducial model, but the flatter IMF favours O enrichment with respect
to model S4.

However, there is another way to obtain an acceptable fit to the
observed abundances. 
In standard chemical evolution models nitrogen production from massive
stars is supposed to be purely secondary, i.e. N is created starting
from seed nuclei of C already present in the gas out of which the
stars were born.
Also low and intermediate mass stars are supposed to produce N in a
secondary fashion but some primary N can be produced in intermediate
mass stars during the third dredge-up episodes in conjunction with
hot-bottom burning (Renzini \& Voli, 1981; Van den Hoek \&
Groenewegen, 1997). 
The consequence of the secondary production is that the abundance of
nitrogen should increase with metallicity in the earliest evolutionary
phases, and that is what most evolutionary models of the Galaxy
predict, not only for the bulge, but also for the solar neighbourhood
(Ballero et al., 2006; Chiappini et al., 2005).

\begin{figure*}
\centering
\includegraphics[width=\textwidth,clip,trim= 0 275 0 0]{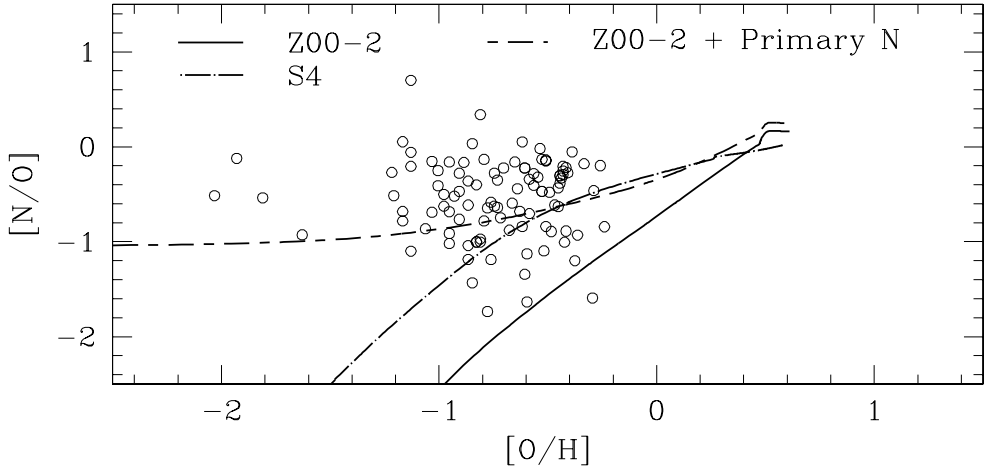}
\caption{Evolution of [N/O] vs. [O/H] (right panels) in the Galactic
  bulge in our fiducial model compared to a model where primary
  production of N from massive stars is assumed (Matteucci,
  1986) and to model S4. Data for N and O in bulge PNe are
  from G\'orny et al. (2004).} 
\label{fig:fig09}
\end{figure*}

Therefore, we investigate what happens to the fiducial model if we
assume that massive stars of all masses produce a constant amount
($0.065 M_{\odot}$) of primary N at every metallicity. 
This hypothesis follows from the heuristic model of Matteucci (1986). 
The results are shown in Fig. \ref{fig:fig09} together with the plot
of model S4.
We see that this model seems to reproduce
the average trend of the observations better than the standard model
and, although it adopts an \emph{ad hoc} assumption, it is useful to
understand that some mechanism of primary production of nitrogen is
likely to occur at any metallicity in massive stars. 
For example, Meynet \& Maeder (2002)
calculated that stellar rotation can produce primary nitrogen in
massive stars, and although Chiappini et al. (2003b) demonstrated that
their rotation yields are insufficient to produce the observed trend
in the solar neighbourhood, this hypothesis is the most promising one,
as shown by Chiappini et al. (2006).
We thus suggest that a continuous primary N production from massive
stars is necessary at any epoch, in analogy with what is required in
the solar vicinity (see Chiappini et al. 2005, for an extensive
discussion on this point).

From Fig. \ref{fig:fig09} it can be also noticed that model S4 differs
from model Z00-2 
with primary N for [O/H]$\lesssim -1$, since while in the latter the
[N/O] ratio is almost independent of [O/H], in the S4 model the plot
curves down sharply at low metallicities due to secondary production
from massive stars, and at [O/H]$\sim -1$ primary production from
low- and intermediate-mass stars becomes dominant.
%It is essential then to acquire measurements of N abundance for [O/H]
%$\leq 1$ (especially if coming from environment which are not expected
%to suffer from self-enrichment), to distinguish whether primary
%production has to be invoked or instead a longer timescale in
%conjunction with a milder star formation is sufficient to explain the
%observed trend (which however, as we showed, is at variance with
 % other constraints).

\section{Summary and conclusions}

In the present paper, we have computed chemical evolution models for
the Galactic bulge where we have considered a scenario involving
evolutionary timescales much faster than in the solar neighbourhood
and in the halo, as the most probable one.

We have adopted new chemical yields suggested by Fran\c cois et
al. (2004) which provided the best fit to the solar neighbourhood
abundance trends.

New observations, unavailable at the time of earlier chemical
evolution models (e.g., Matteucci \& Brocato, 1990; Renzini, 1993;
Matteucci et al., 1999; Moll\'a et al., 2000) have been used to test
these assumptions. 
The agreement between the new observations and our predictions is
quite good, especially for the Fe abundance distribution and the
[$\alpha$/Fe] trends.

Order-of magnitude changes in the main parameters determining chemical
evolution (star formation efficiency, stellar initial mass function,
timescale of gas collapse) were applied in order to explore the
consequences of such changes on the predicted stellar metallicity
distribution and trend of chemical abundances as a function of
metallicity.

%In Tab. \ref{tab:tab1} are briefly reported the results of matching
%the considered models with the observational constraints.

Our main conclusions can be summarized as follows:

\begin{itemize}

\item[-] A short formation timescale combined with a high star
  formation efficiency fits both the observed metallicity distribution
  and chemical abundance ratios. 
  This is typical of a star formation history in a burst regime,
  i.e. strongly concentrated in the first stages of the lifetime of
  the system and vanishing very quickly, in analogy with elliptical
  galaxies.
  We suggest an efficiency of star formation of the order of 20
  Gyr$^{-1}$.
  However the assumption of a closed box must be given up since it
  gives rise to an excessive amount of low-metallicity stars (even
  though not as seriously as in the case of the G-dwarf problem for
  the solar neighbourhood).
  A finite, though small, accretion time is instead required.
  We suggest that this timescale should be of the order of 0.1 Gyr.
 
\item[-] There exists a sort of degeneracy between the gas infall
  timescale and the efficiency of star formation, in the sense that
  values of $\tau$ longer than 0.1 Gyr can combine with values of
  $\nu$ lower than 20 Gyr$^{-1}$ in order to match the observed
    metallicity distribution. However, such a degeneracy is
  broken when we consider also the evolution of [O/Fe] vs. [Fe/H] and
  in any case values of $\tau$ longer than 1 Gyr make it impossible to 
  fit the observed metallicity distribution as well.

\item[-] An IMF flatter than those suitable for the solar neighbourhood
  properties, such as that of Scalo (1998) or Weidner \& Kroupa (2005),
  is necessary to reproduce the observational constraints.
  Namely, a value of $x=0.95$ for massive stars (even smaller than
  that proposed by Matteucci et al., 1999, which was $x=1.1$) gave the
  best overall fit.
  This can be theoretically understood if we note that the star
  formation in the Bulge proceeds like in a burst and there are
  several suggestions in the literature about a top-heavy IMF in 
  starbursts (e.g. Baugh et al. 2005, Nagashima et al. 2005). Figer
  (2005) also finds a flat IMF in the Arches cluster near the Galactic
  center.
  The adopted flattening below $1 M_{\odot}$, as suggested by luminosity
  function measurements (Zoccali et al, 2000) does not affect
  significantly the abundance distribution.
  However, extremely flat IMFs (e.g. that of Z00-1 model),
  if extrapolated to massive stars, 
%can give acceptable results but 
  will lead to some degree of oxygen overproduction.

\item[-] The adoption of the fiducial model explains the
    behaviour of the different [$\alpha$/Fe] abundance ratios with
    metallicity very well, and predicts different slopes for different
    $\alpha-$elements, according to their nucleosynthesis. 
  There is no need to invoke a second infall episode, as 
   suggested by other authors (Moll\'a et al., 2000; Costa et al.,
  2005) to explain the observed values of the abundance ratios.
  We do not exclude however a second infall episode on a much longer
  timescale such as that hypothesized by Moll\'a et al. (2000) as it
  may help explain the presence of a very young stellar population
  confined to the very centre of our Galaxy, but it must involve a
  minor fraction of the bulge gas mass.

\item[-] A certain amount of primary nitrogen from massive stars might
  be required in order to reproduce the average trend of [N/O]
  vs. [O/H] with the fiducial model. 
  The same conclusion was reached for the solar neighbourhood by
  Chiappini et al. (2005) and Ballero et al. (2006).      The
  phenomenon of primary N production from massive stars as a result of
  rotation was studied by Meynet \& Maeder (2002) and Maeder et al
  (2005) and although their yields are still not sufficient to explain
  the observed trend, this seems the most promising way (see Chiappini
  et al. 2006).
  These considerations followed from observations of N and O
  abundances in planetary nebulae; in order to firmly assess this
  point, abundance measurements in stars which have not
  experienced nitrogen self-enrichment are required, especially at low
  metallicities. In alternative, carbon measurements in the same
  planetary nebulae we have considered might allow an estimate of
  the amount of N synthesized by the PN progenitor. 

\item[-] In the near feature new sets of empirical metallicity and
  abundance pattern distributions in different fields and based on
  high resolution spectroscopy will also become available. This will
  dramatically improve the chance of a more robust and quantitative
  comparison between theory and observations, with the ultimate goal
  of drawing the formation and chemical evolution history of the
  Galactic Bulge. 

%  Future observations of the metallicity distribution of bulge
%  stars performed at high resolution, of bulge chemical abundances at
%  very low and high metallicities (especially carbon abundances which
%  are necessary to enlarge our sample) might help put more secure
%  constraints on the details of the models presented here.

\end{itemize}

\small 

\noindent

\subsection*{Acknowledgements}

SKB acknowledges A. Pipino for useful discussions and M. Zoccali for
providing data in a timely manner.

\end{document}